\documentclass[11pt]{article}  

\usepackage{amsmath, amsthm, amssymb}
\usepackage{graphicx}
\usepackage{wrapfig}
\usepackage{sidecap}
\usepackage{floatrow}

\usepackage{authblk}
\usepackage{setspace}
\usepackage{lineno}

\usepackage{apacite}
\usepackage{color}
\definecolor{gray}{gray}{0.5}

\newcommand{\Perr}{e}

\usepackage{lineno}

\title{
A model of reward-modulated motor learning with parallelcortical and basal ganglia pathways
}

\date{}
\begin{document}

\begin{flushleft}

{\Large
\textbf{A model of reward-modulated motor learning with parallel cortical and basal ganglia pathways}
}

\vspace{0.75 cm}

Ryan Pyle$^1,^*$ and 
Robert Rosenbaum$^{1,2,*}$
\\


{\footnotesize
{1}: Department of Applied and Computational Mathematics and Statistics, University of Notre Dame, Notre Dame, Indiana, USA.\\

{2}:Interdisciplinary Center for Network Science and Applications, University of Notre Dame, Notre Dame, Indiana, USA.

* rpyle1@nd.edu, Robert.Rosenbaum@nd.edu
}

\section*{Acknowledgments} 
    This work was supported by National Science Foundation grants DMS-1517828, DMS-1654268, and DBI-1707400. The authors thank Jonathan Rubin, Robert Turner, and Robert Mendez for helpful discussions.

\end{flushleft}



\section*{Abstract}

Reservoir computing is a biologically inspired class of learning algorithms in which the intrinsic dynamics of a recurrent neural network are mined to produce target time series. Most existing reservoir computing algorithms rely on fully supervised learning rules, which require access to an exact copy of the target response, greatly reducing the utility of the system. Reinforcement learning rules have been developed for reservoir computing, but we find that they fail to converge on complex motor tasks.  Current theories of biological motor learning pose that early learning is controlled by dopamine modulated plasticity in the basal ganglia that trains parallel cortical pathways through unsupervised plasticity as a motor task becomes well-learned. We developed a novel learning algorithm for reservoir computing that models the interaction between reinforcement and unsupervised learning observed in experiments. This novel learning algorithm converges on simulated motor tasks on which previous reservoir computing algorithms fail, and reproduces experimental findings that relate Parkinson's disease and its treatments to motor learning. Hence, incorporating biological theories of motor learning improves the effectiveness and biological relevance of reservoir computing models. 

\newpage

~
\vspace{-.75in}
\section*{Introduction}

    Even simple motor tasks require intricate, dynamical patterns of muscle activations. Understanding how the brain generates this intricate motor output is a central problem in neuroscience that can inform the development of brain-machine interfaces, treatments for motor diseases, and control algorithms for robotics. Recent work is largely divided into addressing two distinct questions: How are motor responses encoded and how are they learned? 
    
    From the coding perspective, it has been shown that the firing rates of cortical neurons exhibit intricate dynamics that do not always code for specific stimulus or movement parameters~\cite{Churchland2012,russo2018motor}. A prevailing theory poses that these firing rate patterns are part of an underlying dynamical system that serves as a high-dimensional ``reservoir'' of dynamics from which motor output signals are distilled~\cite{Shenoy2013,Sussillo2014}. This notion can be formalized by reservoir computing models, in which a chaotic or near chaotic, recurrent neural network serves as a reservoir of firing rate dynamics and synaptic readout weights are trained to produce target time series~\cite{Maass2002,Jaeger2004,sussillo2009generating,Lukosevicius2012,Sussillo2014}.

    Reservoir computing models can learn to generate intricate dynamical responses and naturally produce firing rate dynamics that are strikingly similar to those of cortical neurons~\cite{Sussillo2013,Mante2013,Laje2013,Hennequin2014}. However, most reservoir computing models rely on biologically unrealistic, fully supervised learning rules. Specifically, they must learn from a teacher signal that can already generate the target output. Many motor tasks are not learned in an environment in which such a teacher signal is available. Instead, motor learning is at least partly realized through reward-modulated, reinforcement learning rules~\cite{Izawa2011}.

    A large body of studies are committed to understanding how reinforcement learning is implemented in the motor systems of mammals and songbirds~\cite{Brainard2002,Olveczky2005,Kao2005,ashby2010cortical,Izawa2011,Fee2014}. The basal ganglia and their homolog in songbirds play a critical role in reinforcement learning of motor tasks through dopamine-modulated plasticity at corticostriatal synapses. This notion inspired the development of a reward-modulated learning rule for reservoir computing~\cite{hoerzer2012emergence}. However, we found that this learning rule fails to converge on many simulated motor tasks.

    We propose that the shortcomings of previous reservoir computing models can be resolved by a closer inspection of the literature on biological motor learning. 
    A large body of evidence across multiple species supports a theory of learning in which dopamine-modulated plasticity in the basal ganglia or its homologs is responsible for early learning and this pathway gradually trains a parallel cortical pathway that takes over as tasks become well learned or ``automatized''~\cite{Bottjer1984,Carelli1997,Brainard2000,Pasupathy2005,Ashby2007,Obeso2009,Andalman2009,ashby2010cortical,Turner2010,fee2011hypothesis,Olveczky2011}, although the biology is not settled ~\cite{kawai2015motor}. This model of motor learning has only been tested computationally in discrete choice tasks that do not capture the intricate, dynamical nature of motor responses~\cite{Ashby2007}.

    Inspired by this theory of automaticity from parallel pathways, we derived a new architecture and learning rule for reservoir computing. In this model, a reward-modulated pathway is responsible for early learning and serves as a teacher signal for a parallel pathway that takes over the production of motor output as the task becomes well-learned. This algorithm is applicable to a large class of motor learning tasks to which fully supervised learning models cannot be applied and it  outperforms  previous reward-modulated models. We additionally show that our model naturally produces experimental and clinical findings that relate Parkinson's disease and its treatment to motor learning~\cite{Ashby2007,ashby2010cortical,Turner2010}.
    
   \section*{Results}

    We first review two previous learning rules for reservoir computing, then introduce a new, biologically inspired learning rule that combines their strengths.


    \subsection*{FORCE learning}
    
\begin{figure}
    \centering{
    \includegraphics[width = 5in]{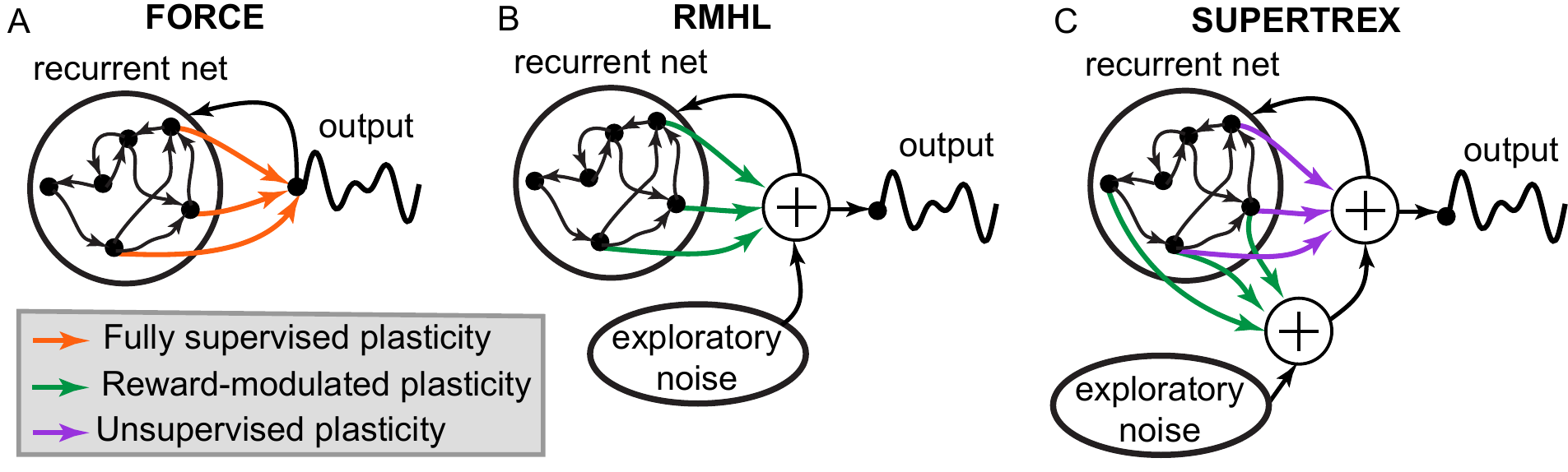}
    \caption{{\bf Network diagrams for three reservoir computing algorithms.} {\bf A)} In FORCE learning,  readout weights are trained to match a target using a fully supervised error signal. Output is fed back to the reservoir. {\bf B)}  RMHL is similar to FORCE, but learning is driven by a reward-modulated plasticity and exploratory noise. {\bf C)} SUPERTREX combines elements of FORCE and RMHL. The exploratory pathway (green) is driven by noise, and acts similarly to RMHL. The mastery pathway (purple) is analogous to FORCE, but uses the output of the exploratory pathway in place of the fully supervised signal used by FORCE. The sum of both pathways provides the total output.
    }
    \label{F:ResDiagrams}
    }
    \end{figure}

    One of the most powerful and widely used reservoir computing algorithms is first-order reduced and controlled error or FORCE~\cite{sussillo2009generating}, which is able to rapidly and accurately learn to generate complex, dynamical outputs. The standard architecture for FORCE is schematized in Fig.~\ref{F:ResDiagrams}A (FORCE variants exist, although the underlying principle is the same). The reservoir is composed of a recurrently connected population of ``rate-model'' neurons. The output of the reservoir is trained to produce a target time series by modifying a set of readout weights, and the output affects the reservoir through a feedback loop.

    The reservoir dynamics are defined by 
    \begin{equation} \label{eq:1}
    \tau \frac{d\textbf{x}}{dt}= -\textbf{x} + J\textbf{r}+ Q \textbf{z}\end{equation}
    Here, 
    \[
    \textbf{r}=\tanh(\textbf{x})  + \boldsymbol{\epsilon} 
    \] 
    is a time-dependent vector representing the activity of units within the reservoir, $\tau$ is a time constant, $\boldsymbol{\epsilon}$ is a small noise term,
    \[
    \textbf{z} = W \textbf{r}
    \]
    is the reservoir output, $J$ is the recurrent connectivity matrix, $Q$ the feedback weights and $W$ the readout weights. A feedforward input term is also commonly included~\cite{sussillo2009generating}, but we do not use one here. 
    When the spectral radius of $J$ is sufficiently large, the intrinsic dynamics of ${\bf r}(t)$ become rich and chaotic~\cite{Sompolinsky1988}. The goal of FORCE  is to utilize these rich dynamics by modifying readout weights, $W$, in such a way that the output, ${\bf z}(t)$, matches a desired target function, ${\bf f}(t)$. 
    A powerful and widely used learning algorithm for FORCE, recursive least squares (RLS), is defined by~\cite{sussillo2009generating}
    \begin{equation}\label{E:dWFORCE}
    \tau_w\frac{dW}{dt} = 
    -\textbf{e}\textbf{r}^TP
    \end{equation}
    where 
    \begin{equation}\label{E:ErrForce}
    \textbf{e}(t)=\textbf{z}(t)-\textbf{f}(t)
    \end{equation}
    is the error vector and $\tau_w$ is the learning time scale. 
    The matrix, $P$, is a running estimate of the inverse of the correlation matrix of rates, $\bf r$ (see Materials and Methods).

    FORCE excels at generating a target time series by harvesting reservoir dynamics, but is incomplete as a model of motor learning. As a fully supervised learning rule, FORCE must have access to the correct output to determine its error (see the presence of $\textbf{f}$ in Eq.~\eqref{E:ErrForce}). Since the correct output must already be generated to compute the error, FORCE can only learn target functions that are already known explicitly and can be generated. Many motor learning tasks require the generation of an unknown target using a lower-dimensional error signal~\cite{Izawa2011}. We consider examples of such tasks below.
    A potential solution for these issues is provided by appealing to biological motor learning, which is controlled at least in part by  dopamine-modulated reinforcement learning in the basal ganglia~\cite{Turner2010,ashby2010cortical,Izawa2011}.

    \subsection*{Reward modulated Hebbian learning}
    Reward-modulated Hebbian learning (RMHL)~\cite{hoerzer2012emergence} is a reinforcement learning rule for reservoir computing in which reward is indicated by a one-dimensional error signal using a plasticity rule inspired by  dopamine-dependent Hebbian plasticity observed in the basal ganglia. RMHL uses the same reservoir dynamics (Eq.~\eqref{eq:1}) and the same basic architecture as FORCE (Fig.~\ref{F:ResDiagrams}B), but the learning rule is fundamentally different. 
    
    The original RMHL algorithm~\cite{hoerzer2012emergence} used a binary error signal, despite potentially poorer performance than other options, to demonstrate that the algorithm could learn with minimal information. 
    We implement a modified version of RMHL with  an error signal, 
    $$
    \Perr({\bf z}(t),t),
    $$ 
    that can be any time-dependent, non-negative function of the output, ${\bf z}$, which the algorithm will seek to minimize. Equivalently, $\Perr$ is proportional to the negative of reward.
    
    In contrast to the fully supervised error signal, ${\bf e}$, used in FORCE learning, $\Perr$ is scalar (one-dimensional) even when ${\bf z}$ is a higher-dimensional vector. Moreover, $\Perr$ can quantify any notion of ``error'' or ``cost'' associated with the output, ${\bf z}$, and does not assume that a target output is known or even that there exists a unique target output. 
    This allows RMHL to be applied to a large class of learning tasks to which FORCE cannot be applied, as we demonstrate below.

    

    
    To decrease error, RMHL makes random perturbations to the reservoir output as a form of exploration. Specifically, the output, $\bf z$, is given by 
    \[
    \textbf{z} = W \textbf{r} + \Psi(\Perr)\boldsymbol{\eta}
    \]
    where $\boldsymbol{\eta}(t)$ is exploratory noise and $\Psi$ is a sublinear function that serves to damp runaway oscillations during learning. 
    The learning rule is then given by 
\begin{equation}\label{E:dWRMHL}
    \tau_w\frac{dW}{dt} = \Phi(
    \hat \Perr)\hat{\textbf z}{\bf r}^T
  \end{equation}
    where $\hat{\textbf x}$ denotes a high-pass filtered version of $\textbf{x}$, which represents recent changes in $\bf x$ and $\Phi$ is a sublinear function that controls when to update the weights. 
    We assume that $\Psi$ is an increasing function and $\Phi$ an odd function with $\Psi(0)=\Phi(0)=0$. This assures that exploration and learning are effectively quenched when the error is consistently near zero. 
    Intuitively, the learning rule can be understood as follows: If a random perturbation from ${\boldsymbol \eta}$ has recently decreased $\Perr$, this perturbation is then incorporated into $W$.




    
    \subsection*{SUPERTREX: A new learning algorithm for reservoir computing}
    
    Unfortunately, on many tasks, the weights trained by RMHL fail to converge to an accurate solution, as we show below. RMHL models dopamine-modulated learning in the basal ganglia, but does not account for experimental evidence for the eventual independence of well-learned tasks on the activity of the basal ganglia: 
    It has been proposed that the basal ganglia are responsible for early learning, but train a parallel cortical pathway that gradually takes over the generation of output as tasks become well-learned and ``automatized''~\cite{Pasupathy2005,Ashby2007,ashby2010cortical,Turner2010,Helie2012}. This could explain why some neurons in the basal ganglia are active during early learning and exploration, but inactive as the task becomes well-learned~\cite{Carelli1997,miyachi2002differential,Pasupathy2005,poldrack2005neural,Ashby2007,Tang2009,ashby2010cortical,Helie2012}. It could also explain why patients or animals with basal ganglia lesions can perform previously learned tasks well, but suffer impairments at learning new tasks~\cite{miyachi1997differential,Obeso2009,Turner2010}. This idea is also consistent with many findings suggesting that the basal ganglia homolog in song birds is responsible for early learning and exploration of novel song production, but not for the vocalization of well-learned songs~\cite{brainard2004contributions,Kao2005,Aronov2008,Andalman2009,fee2011hypothesis}.

    The FORCE and RMHL algorithms could be seen as analogous to the individual pathways in this theory of motor learning: RMHL learns through reward-modulated exploration analogous to the basal ganglia, while FORCE models cortical pathways that learn from the output produced by the basal ganglia pathway. Inspired by this analogy, we next introduce a new algorithm, Supervised Learning Trained by Rewarded Exploration (SUPERTREX), that combines the strengths RMHL and FORCE to overcome the limitations of each. 
    
    The architecture of SUPERTREX (Fig.~\ref{F:ResDiagrams}C) is different than the architectures of FORCE and RMHL: There are now two distinct sets of weights from the reservoir to the outputs, and each is trained with a separate learning rule. The ``exploratory pathway'' learns via an RMHL-like, reinforcement learning algorithm, requiring only a one-dimensional metric of performance rather than an explicit error signal. The exploratory pathway is roughly based off of the biological basal ganglia pathway. 
    The ``mastery pathway'' learns through a FORCE-like algorithm. The key idea is that the activity of the exploratory pathway can act as a target for the mastery pathway to learn, replacing the supervised error signal required by FORCE. Hence, SUPERTREX does not need the explicit supervisory error signal that FORCE does. The mastery pathway is roughly based off of the biological cortical pathway.
    
    
    Importantly, the convergence issues we have found with RMHL are not problematic for SUPERTREX because weights in the RMHL-like exploratory pathway do not need to converge to a correct solution because weights in the mastery pathway converge instead.
    
    
    SUPERTREX uses the same reservoir dynamics (Eqs.~\eqref{eq:1}), but the outputs are  determined by 
    \[
    \begin{aligned}
    &{\bf z}_1 = W_1 \textbf{r} + \Psi(\Perr)\boldsymbol{\eta} \\
    &{\bf z}_2 = W_2 \textbf{r} \\
    &\textbf{z} = {\bf z}_1 + {\bf z}_2.
    \end{aligned}
    \]
    Here, ${\bf z}_1$ is the output from the exploratory pathway, ${\bf z}_2$ from the mastery pathway, and $\bf z$ is the total output. 
    The learning rules are defined by
\begin{equation}\label{E:dWST}
    \begin{aligned}
    &\tau_{w1}\frac{dW_1}{dt} = \Phi(\hat \Perr)\hat{\textbf z}\,{\bf r}^T \\
    &\tau_{w2}\frac{dW_2}{dt} = ({\bf z}-{\bf z}_2){\bf r}^TP.
    \end{aligned}
  \end{equation}
    Intuitively, the first learning rule works like RMHL to quickly minimize the total error, as it uses the error of the total output, $\textbf{z}$.  However, it only controls the ${\bf z}_1$ component of $\textbf{z}$, resulting in ${\bf z}_1 + {\bf z}_2 \approx \textbf{f}$. Error between the ${\bf z}_2$ component and $\textbf{f}$ is therefore just ${\bf z}_1$, which replaces the error in the second learning rule since ${\bf z}-{\bf z}_2={\bf z}_1$. As ${\bf z}_2$ approaches $\textbf{f}$, learning in the exploratory pathway causes ${\bf z}_1$ to approach $\bf 0$ in order to keep the total $\textbf{z}$ correct. 
    
    Additionally, we added one extra component to the SUPERTREX algorithm. Learning transfer from the exploratory pathway is soft thresholded based on total error - if error grows above this point, the transfer rate is gradually reduced to 0. This means that transfer can only occur if the total combined output of both pathways is correct. In practice, this is true for the entire learning period except for a small initial period while the exploratory pathway is finding a solution. Performance without this addition was similar overall, but slightly slower.
    
    Note that the learning rule for $W_1$ is local in the sense that it only involves values of the presynaptic and postsynaptic  variables in addition to the error signal, $\Perr$. The learning rule for $W_2$ would be local if it were not for the computation of $P$, which is biologically unrealistic. However, $P$ can be replaced by the identity matrix to make the learning rules for SUPERTREX purely local.  This slows down learning, but the network can still learn to produce target outputs from a one-dimensional error signal (see \cite{hoerzer2012emergence} and the disrupted learning example below).
    
    In summary, RMHL-like learning in the exploratory pathway uses a one-dimensional error signal, $\Perr$, to track the target while FORCE-like learning in the mastery pathway uses the exploratory pathway as a teacher signal until it learns the output and takes over. This models current theories of biological motor learning in which early learning is dominated by dopamine-dependent plasticity in the basal ganglia, which gradually trains parallel cortical pathways as the task becomes well-learned.

    We next test SUPERTREX on three increasingly difficult motor tasks, comparing the performance of SUPERTREX to that of FORCE and RMHL.


    \subsection*{Task 1: Generating a known target output}
    
\begin{figure}
    \centering{
    \includegraphics[width=3in]{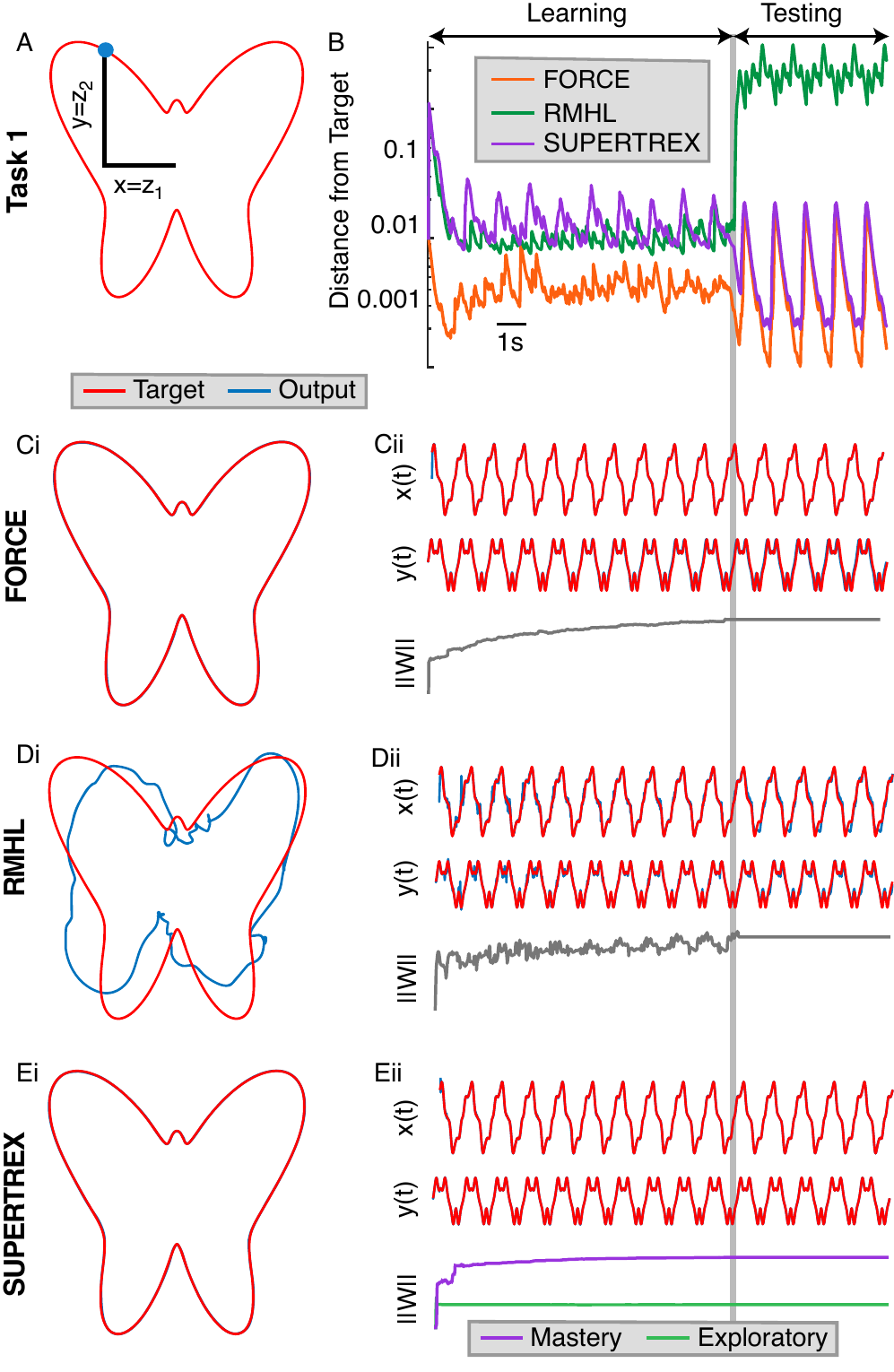}
    \caption{{\bf Performance of three learning algorithms on Task 1.} 
    {\bf A)} Task 1 is to draw a butterfly curve by directly controlling the $x$- and $y$-coordinates of a pen. Specifically, the outputs of the reservoir is ${\bf z}(t)=(x(t),y(t))$ where $x(t)$ and $y(t)$ are the Cartesian coordinates of the pen. {\bf B)} Euclidean distance of pen from target for FORCE (orange),  RMHL (green), and SUPERTREX (purple). Learning was halted by freezing weights and exploration after ten periods, so the remaining five periods represent a testing phase. 
    {\bf Ci)} Target butterfly curve (red) versus the butterfly drawn by FORCE (blue) during the testing phase. {\bf Cii)} Target (red) and actual (blue) outputs, $x(t)$ and $y(t)$, and the norm of the weight matrix, $\|(WW^T)^{\frac{1}{2}}\|_2$, produced by FORCE. {\bf Di-ii)} Same as C, but for RMHL. {\bf Di-ii)} Same as C, but for SUPERTREX and the norm of each matrix, $W_1$ (exploratory; green) and $W_2$ (mastery; purple) are plotted separately. Vertical gray bar indicates the time at which weights were frozen. Note exploratory weight do change, but primarily at the start and is hard to see over the full trial timescale. See figure~\ref{F:Swap} for more details.
    }
    \label{F:Task1}
    }
    \end{figure}

    We first consider a task in which the goal is to draw a parameterized curve of a butterfly by directly controlling the coordinates of a pen (Fig.~\ref{F:Task1}A). Specifically, the target is given by ${\bf f}(t)=(x(t),y(t))$ where $x(t)$ and $y(t)$ parameterize the $x$- and $y$-coordinates of a pen that successfully traces out the butterfly. The reservoir output, ${\bf z}(t)$, controls the coordinates of the pen, so the goal is to train the weights so that ${\bf z}(t)$ closely matches the target, ${\bf f}(t)$.

    The learning algorithms are first allowed to learn for ten repetitions of the task. As a diagnostic, the error signals are not computed and the 
    weights are frozen for a further five repetitions. This provides a way to check the accuracy of the final solution, demonstrating whether or not the algorithm has converged to an accurate solution. During this testing phase, feedback to the system comes from the true solution~\cite{sussillo2009generating}. Specifically, $Q{\bf z}$ is replaced by $Q{\bf f}$ in Eq.~\eqref{eq:1}. This avoids a drift in the phase of the solution that otherwise occurs when weights are frozen (see below). 
    Additionally, for SUPERTREX, the exploratory pathway was shut off during these last five repetitions (${\bf z}_1$ set to zero) to test how well the mastery pathway converged.

    This simple task is well suited to FORCE, which requires a known target, ${\bf f}$, in order to compute the fully supervised error signal, 
    \[{\bf e}={\bf z}-{\bf f}.\]
    FORCE was able to quickly find the correct solution to the task and maintained the correct result even after weights were frozen (Fig.~\ref{F:Task1}B,C). Another measure of convergence is the activity of weight matrix, $W$, which quickly converged then stabilized (Fig.~\ref{F:Task1}Cii, bottom). Error  also remained low after learning was disabled (Fig.~\ref{F:Task1}B). In summary, as expected, FORCE learned this task quickly and accurately.

    %
    
    
    To apply RMHL and SUPERTREX to this task, we set 
    \[
    e=\|{\bf z}-{\bf f}\|^2
    \]
    where $\|\cdot \|$ is the Euclidean norm, \emph{i.e.} the distance of the pen from its target. This error contains strictly less information than the fully supervised error used by FORCE. RMHL performed well during learning, but the performance after weights were frozen (Fig.~\ref{F:Task1}B,D) along with the cyclical changes in $\|W\|$ during learning (Fig.~\ref{F:Task1}Dii) demonstrate that the RMHL algorithm never actually converged. Instead, RMHL relied on rapid  changes in $W$ to mimic the correct output at each time point without  truly learning it. Even when the number of learning trials was dramatically increased (to 100, not shown), RMHL's $W$ continually oscillated rather than converging.

    SUPERTREX performed well on this task. During learning SUPERTREX performed slightly worse than FORCE and similar to RMHL (Fig.~\ref{F:Task1}B,E). Unlike RMHL, though, SUPERTREX continued to track the target after learning was disabled, and performed similarly to FORCE during that phase (Fig.~\ref{F:Task1}B). This, combined with the apparent convergence of $\|W\|$ during learning (Fig.~\ref{F:Task1}Eii, purple and green curves converge), indicates that the SUPERTREX algorithm did converge - albeit more slowly than FORCE.

    Interestingly, SUPERTREX produced less error during the testing phase than during learning (Fig.~\ref{F:Task1}B). This is because exploration introduces random errors during learning, but exploration was turned off during testing so that output was produced only by the well-trained mastery pathway. This is comparable to findings in song birds in which  natural or artificial suppression of neural activity in brain areas homologous to the basal ganglia reduce exploratory song variability and vocal errors~\cite{Kao2005}.
    
\begin{figure}
    \centering{
    \includegraphics[scale = .4]{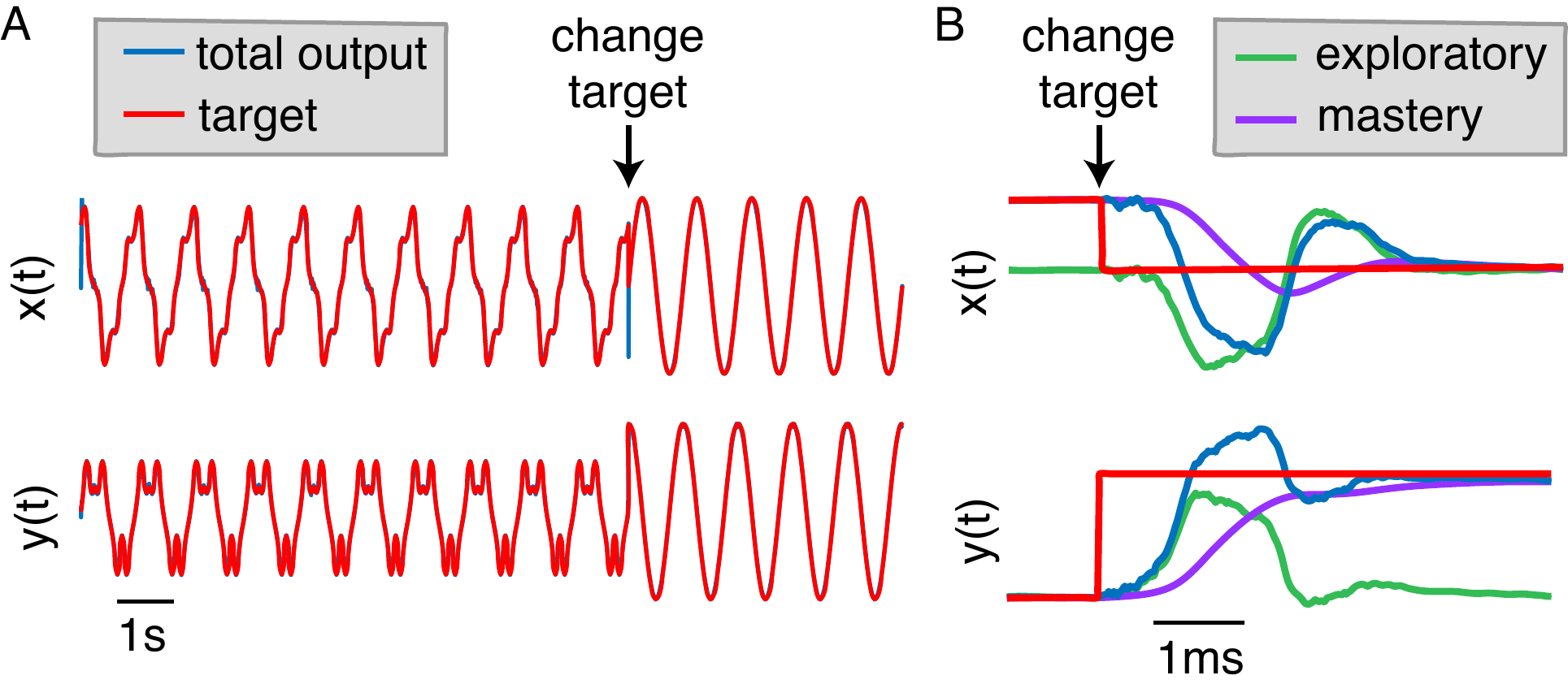}
    \caption{{\bf The dynamics of SUPERTREX with an abruptly changed target.} {\bf A)} Target (red) and actual (blue) output. Same as Task 1, but the target was changed from a butterfly to a circle after ten periods.  {\bf B)} Detail around the time of change. Same colors as A (red for target and blue for total output), with addition of exploratory (green) and mastery (purple) components of the total output. Note that exploratory + mastery = total output. }
    \label{F:Swap}
    }
    \end{figure}

    Looking at Fig.~\ref{F:Task1}, it can be hard to tell whether the exploratory pathway is active, since the weights do not seem to change. This is due to the large timescale of the trial compared to the exploratory dominated phase, which only occurs as the algorithm is first adjusting to the task. An interesting illustration of the exploration / mastery handoff in SUPERTREX is provided by suddenly changing the target from a butterfly to a circle during learning (Fig.~\ref{F:Swap}). The relative contributions from the exploratory pathway and the mastery pathway show that the exploratory pathway initially tracks the new target (Fig.~\ref{F:Swap}B). Since the exploratory pathway is equivalent to RMHL, we know from above that the pathway is only mimicking the output through rapid weight changes. Over time, the mastery pathway learns from the activity of the exploratory pathway and begins taking over the generation of the output. This ``handoff'' from the exploratory to the mastery pathway produces a damped oscillation around the target (Fig.~\ref{F:Swap}B).

    
    \subsection*{Task 2: Generating an unknown target from a scalar error signal}
    
    Task 1 is a simple introductory task to compare the three learning algorithms, but it is also  unrealistic in some ways that play towards FORCE's strengths. Specifically, the task involves producing an output, $\bf z$, to match a known target, $\bf f$, and error is computed in terms of the difference between $\bf z$ and $\bf f$. 
    In many tasks, motor output has indirect effects on the environment and the target and error are given in terms of these indirect effects.  For example, consider a human or robot performing a drawing task. Motor output does not control the position of the pen directly, but instead controls the angles of the arm joints, which are nonlinearly related to pen position.   On the other hand, error might be evaluated in terms of the distance of the pen from its target. 
    Task 2 models this scenario. 
    
    

\begin{figure}
    \centering{
    \includegraphics[width = 3in]{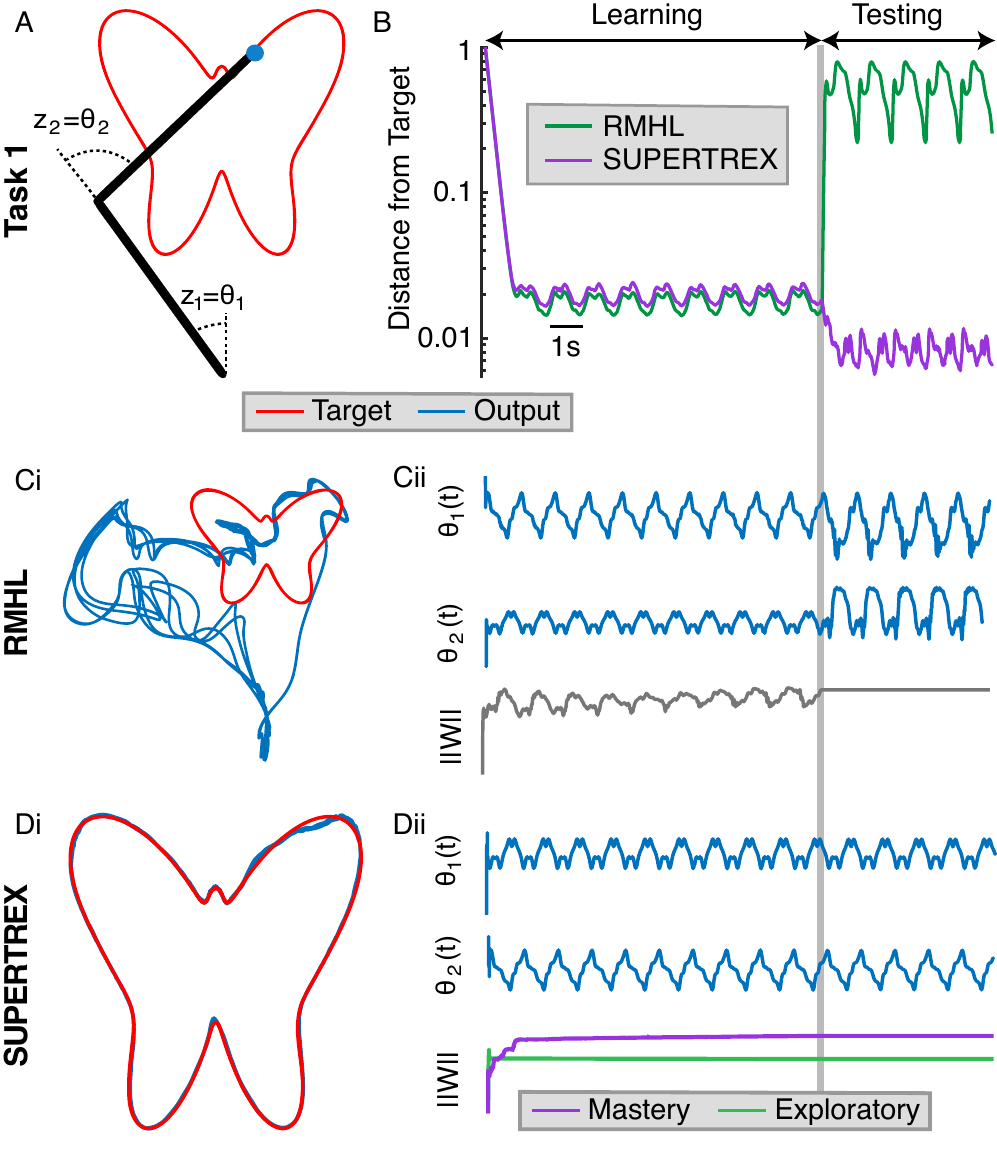}
    \caption{{\bf Performance of RMHL and SUPERTREX on Task 2.} {\bf A)} Task 2 is to draw the same butterfly (red) as in Task 1, but the reservoir output now controls the arm joint angles, $z_1(t)=\theta_1(t)$ and $z_2(t)=\theta_2(t)$. Error is still computed in terms of pen coordinates. FORCE is not applicable to this task. {\bf B)} Euclidean distance of pen from target for RMHL (green) and SUPERTREX (purple). Learning was halted by freezing weights after ten periods, so the remaining five periods represent a testing phase. {\bf C-D)} Same as Fig.~\ref{F:Task1}D-E except that angles, $\theta_1(t)$ and $\theta_2(t)$ are plotted in place of pen coordinates.  
    }\label{F:Task2}
    }
    \end{figure}

    The goal in Task 2 is to draw the same butterfly from Task 1, parameterized by the same target coordinates ${\bf f}(t)=(x(t),y(t))$. However, the reservoir output controls the angles of  two arm joints (Fig.~\ref{F:Task2}A),
    \[
    {\bf z}(t)=(\theta_1(t),\theta_2(t)).
    \] 
    We assume that the subject does not have access to the target angles that draw the butterfly. Instead, they only have access to the target pen coordinates, ${\bf f}$, and its distance to the actual pen coordinates, which are related to the angles through a nonlinear function,
    \[
    h(\theta_1,\theta_2)=(x,y).
    \]
    FORCE cannot be applied directly to this task since the fully supervised error required for FORCE would need to be computed in terms of target angles instead of target pen position.
    
    RMHL and SUPERTREX can be applied to this task since they only require a signal that provides enough information to determine whether error recently increased or decreased. 
    In particular, this is accomplished by setting
    \[
    \Perr=\|h({\bf z})-{\bf f}\|^2
    \]
    where $h({\bf z})=(x,y)$ is the pen position.


    Once again, the task is divided into 10 learning cycles and 5 test cycles, with learning algorithms and the exploratory pathway of SUPERTREX disabled during the test cycles. Since the target angles, ${\bf z}(t)$, are unknown, feedback during testing cannot be replaced by the target as was done for Task 1. Instead, it is provided by the output from five previous periods.

    
    
    RMHL performed poorly on this task. It eventually  mimicked the target (Fig.~\ref{F:Task2}B), but once again failed to converge  (Fig.~\ref{F:Task2}B,C). 
    SUPERTREX was able to track the target, and continue to produce it even after weight changes ceased (Fig.~\ref{F:Task2}B,D). 
    Hence, the combination of FORCE-like learning and RMHL-like learning implemented by SUPERTREX is able to learn a task that neither FORCE nor RMHL can learn on their own.

    

    \subsection*{Task 3: Learning and optimizing a task with multiple candidate solutions}
    
    While FORCE cannot be applied to Task 2  as it is currently defined, it could be applied if the inverse of $h$ were explicitly computed off-line to provide the target angles, $(\theta_1,\theta_2)=h^{-1}({\bf f})$, from which to compute a fully supervised error signal. This approach assumes that the subject knows the inverse of $h$ and therefore does not easily extend to learning tasks in which $h$ is difficult or impossible to invert. We now consider a task in which the error is not an invertible function of the motor output.

\begin{figure}
    \centering{
    \includegraphics[width = 3in]{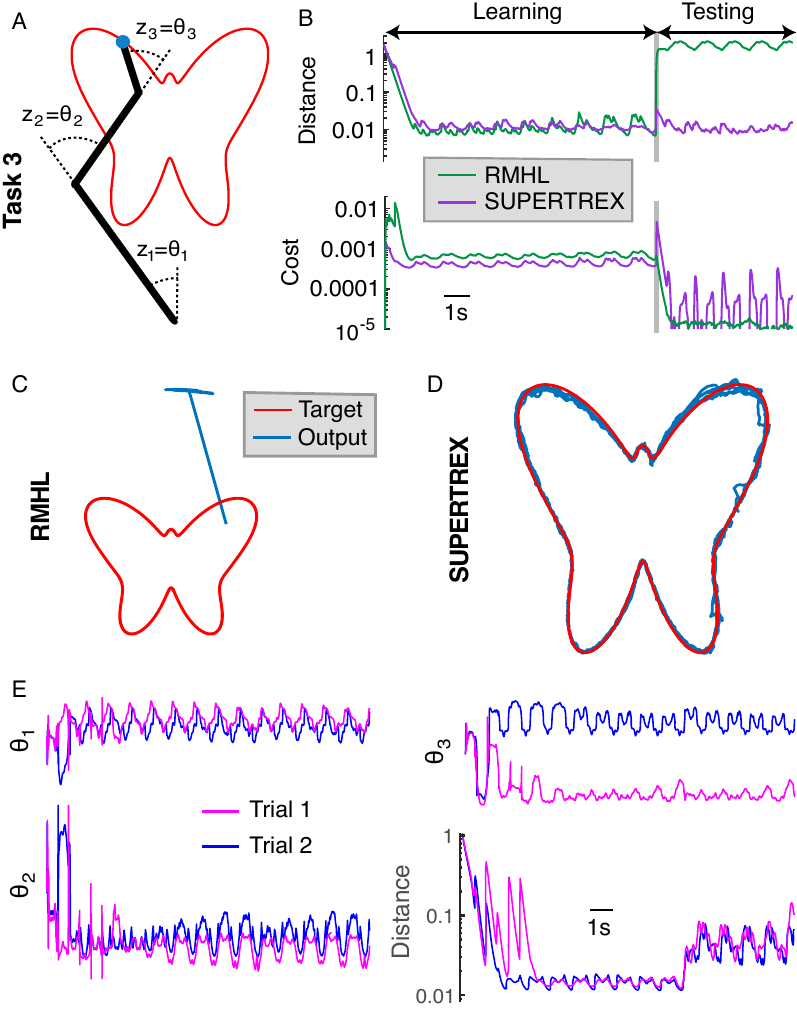}
    \caption{{\bf Performance of RMHL and SUPERTREX on Task 3.} {\bf A)} Task 3 is to draw the same butterfly (red) as in Tasks 1 and 2, but the reservoir output now controls three arm joint angles, $z_1(t)=\theta_1(t)$, $z_2(t)=\theta_2(t)$, and $z_3(t)=\theta_3(t)$ with a different cost function associated to moving each joint. Error is computed in terms of pen coordinates and cost of moving joints. FORCE is not applicable to this task. {\bf B)} Euclidean distance of pen from target (top) and cost (bottom; $C(\theta_1',\theta_2',\theta_3')$) for RMHL (green) and SUPERTREX (purple). 
    {\bf C,D)} Same as Di and Ei in Fig.~\ref{F:Task1}. \bf{E)} Angular outputs and distance from target across two different SUPERTREX trials. The overall solution found was similar, with a mirrored rotation in one joint angle. 
    }
    \label{F:Task3}
    }
    \end{figure}

    Specifically, we consider an arm with three joints (Fig.~\ref{F:Task3}A) and a cost function, $C( \theta_1', \theta_2', \theta_3')$, that penalizes the movement of some joints more than others. Here, $ \theta_j'$ is the time-derivative of $\theta_j$. SUPERTREX can work with any penalty structure, making the choice arbitrary. Given that, we decided to loosely model our arm on a real human arm, with the joints corresponding to shoulder, elbow, and wrist. The penalties are larger for the angles controlling larger arm lengths, so the cost is lowest for the wrist joint, $\theta_3$, and largest for the shoulder joint, $\theta_1$, based on the fact that you are more likely to move your wrist than your entire shoulder and arm for a small reaching task - this can also be seen as an energy conservation principle, with larger costs associated to the more costly shoulder joint compared to the wrist joint.
    
    For this task, there are infinitely many candidate solutions that successfully draw the butterfly, differing by the cost of joint movement. This turns our learning task into an optimization problem. 
    
    Using FORCE for this task does not make sense, as it would require prior, explicit knowledge of the desired time series of joint angles.  Essentially, it would require that the optimization problem had already been solved offline. 
    RMHL and SUPERTREX can be applied to this problem by setting
    \[
    \Perr=\|h({\bf z})-{\bf f}\|^2+C(\theta_1',\theta_2',\theta_3').
    \]
    In this context, RMHL and SUPERTREX work as greedy search algorithms that make local changes to the angular output to reduce error and cost.  Note, however, that the solution they find may not be globally optimal.

    We applied RMHL and SUPERTREX to this task using the same protocol for the learning and testing phases that we used for Task 2.
    RMHL performed poorly on this task (Fig.~\ref{F:Task3}B,C), which is not surprising considering its poor performance on Task 2. 
    SUPERTREX performed much better than RMHL: It was able to track the target, and continue to produce it even after weights were frozen (Fig.~\ref{F:Task3}B,D). Over multiple runs SUPERTREX will find different solutions, as seen in Fig 5 E). The solution found will primarily depend on the initial condition, but the randomness in searching will also play a role. In this task, SUPERTREX tended to find similar solutions, except for random mirroring of certain angles.
    In summary, SUPERTREX can solve motor learning tasks in which there are multiple ``correct'' solutions with different costs. 

    \subsection*{Disrupted learning  as a model of Parkinson's disease}

    The design of SUPERTREX was motivated in part by observations about the role of the basal ganglia in motor learning and Parkinson's disease (PD). PD is caused by the death of dopamine producing neurons in the basal ganglia, resulting in motor impairment. A common treatment for PD is a lesion of basal ganglia output afferents. Such lesions alleviate PD symptoms and impair performance on new learning tasks more than well-learned tasks~\cite{Obeso2009,Turner2010}. These and other findings have inspired a theory of motor learning in which the basal ganglia are responsible for early learning, but not the performance of well-learned tasks and associations~\cite{Turner2010,Helie2012}. SUPERTREX is consistent with this theory if the exploratory pathway is interpreted as a basal ganglia pathway and the mastery pathway the cortical pathway. To test this model, we next performed an experiment in SUPERTREX that mimics the effects of PD and its treatment with basal ganglia lesion. 
    
    The hand-off of learning from the exploratory to mastery pathway occurs extremely quickly in SUPERTREX due to the powerful, but biologically unrealistic RLS learning rule used in the mastery pathway (see, \emph{e.g.}, Fig.~\ref{F:Swap}). To make SUPERTREX more biologically plausible for this experiment, we replaced the RLS learning rule with a least-mean-squares (LMS) rule by replacing $P$ in Eq.~\eqref{E:dWFORCE} with the identity matrix~\cite{hoerzer2012emergence}. This modified rule is more realistic because it avoids the complicated computation of the matrix, $P$, it makes the learning rules local, and it causes the mastery pathway to learn more slowly, which slows the hand-off from the exploratory pathway.

\begin{figure}
    \centering{
    \includegraphics[width = 4.5in]{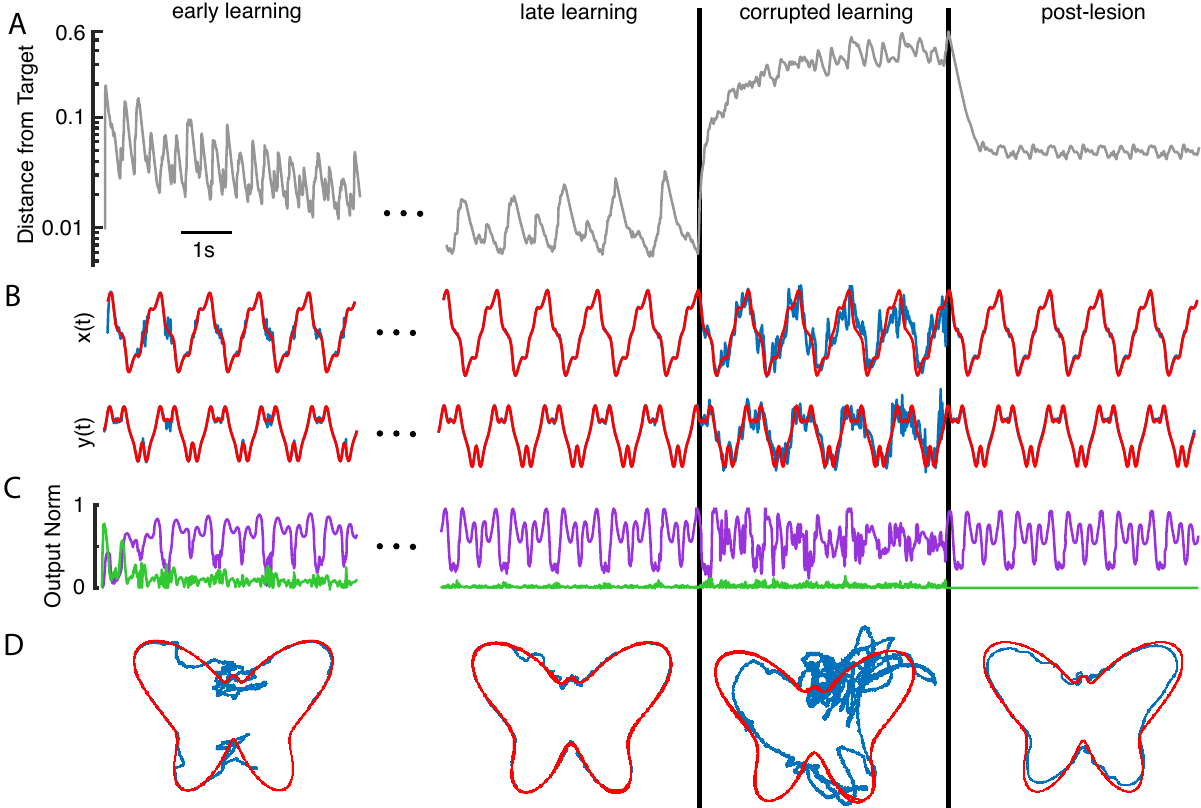}
    \caption{{\bf SUPERTREX with a corrupted error signal models Parkinson's disease and its treatment.} {\bf A)} Euclidean distance of pen from its target for a modified version of SUPERTREX on Task 1. Learning proceeded normally for 100 trials. First five trials (early learning) and last five trials (late learning) are shown. The error signal was corrupted over the following five trials (corrupted learning) and the exploratory pathway was lesioned for the last five trials (post-lesion). {\bf B)} Target (red) and actual (blue) outputs. {\bf C)} Normed outputs from mastery pathway (purple) and exploratory pathway (green). {\bf D)} Target butterfly (red) versus drawn curve during each of the plotted groups of five trials. 
    }
    \label{F:PARK}
    }
    \end{figure}

    We applied this modified SUPERTREX algorithm to Task 1. For 100 trials, learning proceeded normally. SUPERTREX learned the target more slowly than in Fig.~\ref{F:Task1} and with a slight degradation in performance due to the use of LMS instead of RLS learning in the mastery pathway (Fig.~\ref{F:PARK}, early and late learning).  This phase models normal learning before the onset of PD. By the end of this phase (Fig.~\ref{F:PARK}, late learning), the task has become ``well-learned'' in the sense that output is generated by the mastery pathway instead of the exploratory pathway. The system output depending primarily on the mastery rather than exploratory pathway can be seen in Fig~\ref{F:PARK} C).

    For the next five trials, we corrupted the error signal to model the effects of PD. Since $\Perr$ models the error or cost of motor output, it is negatively related to dopamine release. Specifically, $\hat \Perr\propto D_{max}-D$ where $D$ quantifies dopamine release and $D_{max}$ is the maximum possible value of $D$. Hence, PD-induced dopamine depletion is modeled by artificially increasing $\hat \Perr$, which we achieve by setting 
    \[
    \Perr=\|{\bf f}-{\bf z}\|^2
    \]
    \[
    \hat \Perr = \hat \Perr + p
    \]
    where $p(t)$  increases over time. 
    Here, $p= 0$ corresponds to a healthy subject and, as $p$ increases, SUPERTREX falsely evaluates more of its actions as being in error or costly. For our Parkinsonian task, we chose a $p(t)$ that linearly increased to .1 over the duration of the corrupted learning phase.
    
    Even though the mastery pathway had taken over motor output before the error signal of the exploratory pathway was corrupted, the perceived increase in error caused the exploratory pathway to take over during the corrupted learning phase because the contribution of the exploratory pathway increases with error. Although the actual disruption may seem small (See Fig~\ref{F:PARK} C), where the exploratory activity is similar to that of early learning), the mismatch between actual error and perceived error during the corrupted learning phase results in highly inaccurate motor output (Fig.~\ref{F:PARK}, corrupted learning phase) as activity leaves the learned manifold and is unable to recover. These results model the motor impairments associated with PD. Indeed, PD symptoms are believed to be caused, at least in part, by aberrant learning in the basal ganglia~\cite{Turner2010,ashby2010cortical}.
    
    

    

    In the last five trials, we disabled the exploratory pathway, modeling basal ganglia lesion, and the feedback term, $Q{\bf z}$, was replaced by $Q{\bf f}$ in Eq.~\eqref{eq:1} (see below and Discussion). SUPERTREX recovered nearly correct output during this last stage (Fig.~\ref{F:PARK}, post-lesion phase) because the output had been stored in the mastery pathway before learning in the exploratory pathway was corrupted. 
    
    As shown in Fig.~\ref{F:PARK}, immediately before corruption began the mastery pathway was essentially solely responsible for generating the correct output. After the Parkinsonian effect, the final output is given solely by the mastery pathway as the malfunctioning exploratory pathway is lesioned. Thus, any degradation in the drawn butterfly is due to harmful changes made to the mastery pathway during the Parkinsonian effect. There are two main reasons why these harmful changes should be small. One is that the exploratory pathway changes are only kept if they result in a decrease in error \textit{even after taking into account the additional Parkinsonian error}, or that the Parkinsonian error term makes changes due to exploration less likely to be accepted. Additionally, for sufficiently large errors the SUPERTREX component that controls transfer from exploration to mastery pathways shuts down, limiting the degree to which harmful perturbations can be assimilated. Thus, post-lesion performance will depend on the specific $p(t)$ Parkinsonian effect used, along with the overall duration of the Parkinsonian effect.

    \subsection*{State Information Promotes Stability of Learned Output}
    
    During our previous examples comparing FORCE, RMHL, and SUPERTREX, the comparison was made by allowing 10 trials of training the algorithm, and then with 5 trials of the learning algorithm shut off (weights frozen) to see if the method had converged. During this testing phase, feedback was modified. In task 1 it was replaced with the correct output (the target), and for tasks 2 and 3 it was replaced with the output from previous trials during learning, which nearly matched the target due to the learning algorithm being active. This allowed us to check whether an algorithm had converged, in the sense that there would be no further feedback and weight changes. However, providing the correct answer as feedback, also known as teacher forcing, could be considered cheating here. Teacher forcing essentially ignores stability of the solution and instead only checks whether the system can correctly produce the next time step of the solution given a perfect fit to the current time step. In order to address this, we repeated task 1, but without teacher forcing.

\begin{figure}
    \centering{
    \includegraphics[width = 4.5in]{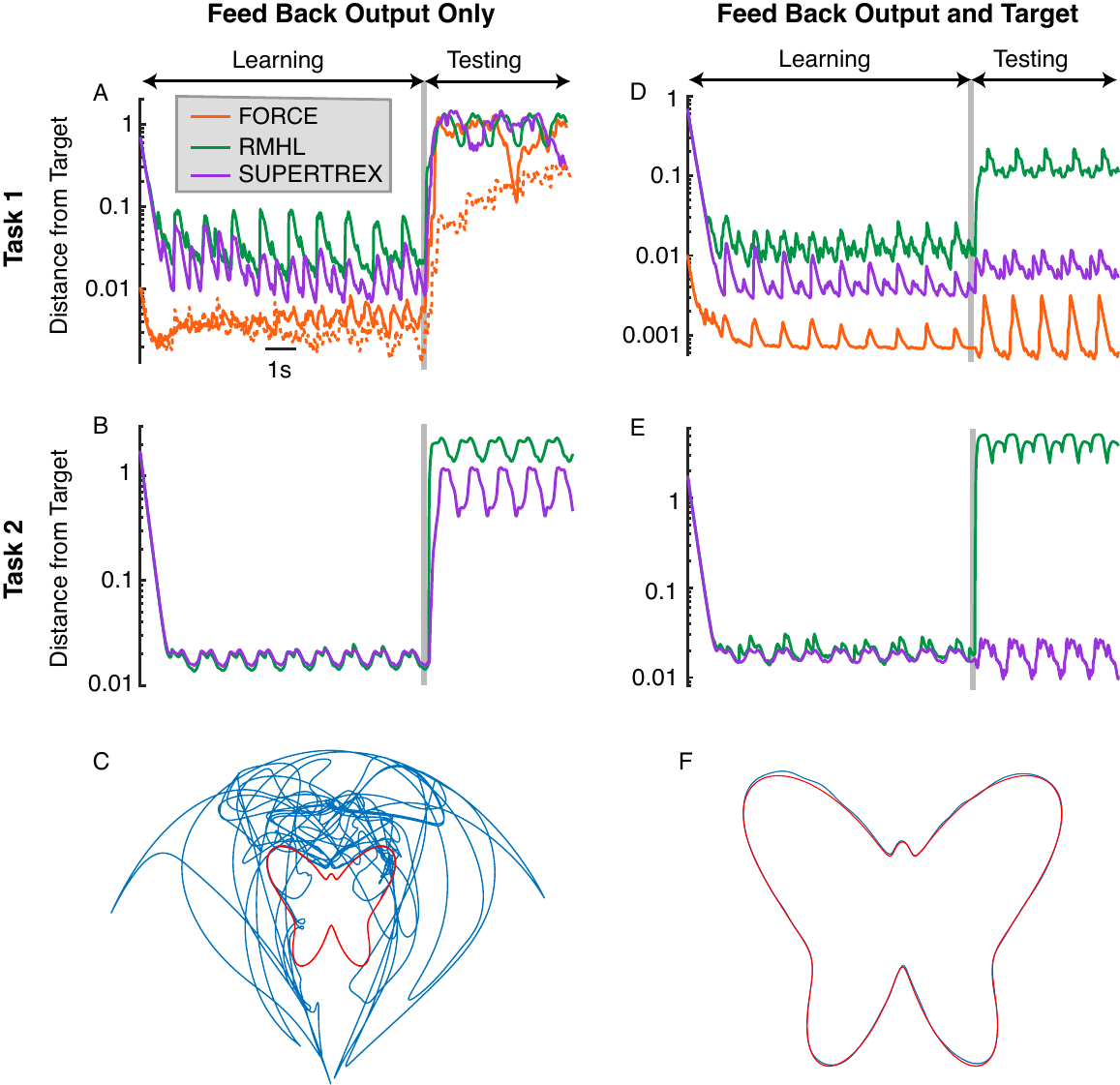}
    \caption{{\bf Including target information in feedback promotes stability without teacher forcing.} {\bf A)} Euclidean distance of pen from its target for FORCE (orange), RHML (green), and SUPERTREX (purple) on Task 1. Same as Fig.~\ref{F:Task1}B except feedback during the testing phase was not replaced with the true solution (teacher forcing), but is instead given by $Q{\mathbf z}$ exactly. {\bf B)} Same as A, but for Task 2 and without FORCE (since it cannot be applied to Task 2). 
    {\bf C)} Butterfly drawn by SUPERTREX from the simulation in B. 
    {\bf D-F)} Same as A-C, except feedback was augmented by the target, $Q[{x\; y\; \mathbf f}]$.
    }
    \label{F:Fdbk}
    }
    \end{figure}

    FORCE has previously been shown to perform well in the absence of teacher forcing~\cite{sussillo2009generating,Abbott2016}, but it failed in our simulations (Fig.~\ref{F:Fdbk}A, solid orange). We suspected that this was due to the extra additive noise, $\boldsymbol \epsilon$, added during learning. Noise is not typically included in applications of FORCE, but reservoir learning is known to be sensitive to noise and other perturbations~\cite{vincent2016driving,Sussillo2014,miconi2017biologically}, which are ubiquitous in biological neuronal networks. Indeed, FORCE performed better when this noise was removed (Fig.~\ref{F:Fdbk}A, dashed orange). 
Noise is an inherent part of RMHL and SUPERTREX, so they cannot be tested without it. Unsurprisingly, RMHL and SUPERTREX also perform poorly without teacher forcing (Fig.~\ref{F:Fdbk}A,B,C). In summary, learning a noisy version of the target prevents all three algorithms from reproducing the target post-learning in the absence of teacher forcing. 
    
%
%

    We resolve this issue by augmenting the feedback to include full information about the state of the system, allowing the system to self-correct. Specifically, we concatenated the $x$- and $y$-coordinates of the target pen position onto the feedback signal, replacing the $Q\mathbf{z}$ term in Eq.~\eqref{eq:1} with $Q[\mathbf{z} \;\; \mathbf{f}]$, during both training and testing. Under this modified framework, we again tested all three algorithms on Task 1 and tested RMHL and SUPERTREX on Task 2. For Task 1, this change is analogous to teacher forcing (since the target coordinates are the same as the target reservoir output). For Task 2, it is distinct from teacher forcing because the  feedback is in terms of the Cartesian coordinates of the target, whereas the output must be in terms of arms' angles. Hence, for Task 2, the system must learn to self correct: If $\mathbf{z}$ and $\mathbf{f}$ differ then the networks need to learn how to generate the correct $\theta_1$ and $\theta_2$ to correct the error. This change greatly improved accuracy of FORCE and SUPERTREX, but not RMHL (Fig.~\ref{F:Fdbk}D,E,F). 

Note that this change is not the same as just providing the correct answer as teacher forcing does. Teacher forcing essentially ``resets'' the system to be correct after every time step by replacing $Q\mathbf{z}$ with $Q\mathbf{f}$, preventing drift. The augmented feedback instead provides sufficient information for the system to be autonomously self-correcting and the feedback is provided as-is with no context. 
In task 2, the algorithm does not have access to the solution it must produce (in terms of arm angles), but only has access to the target pen coordinates, which are non-linearly related to arm angles. This is akin to including a sensory feedback term, where the algorithms have sensory information about the actual and target positions, but do not have explicit information on necessary joint movements to make them overlap. Note that simply replacing the feedback from position to target will not result in convergence, e.g. replacing $Q\mathbf{z}$ with $Q\mathbf{f}$ throughout training and testing does not work. Both pieces of information together are required to build a stable system.
    
 The extra feedback term can be simplified further by changing $\mathbf{f}$ into a simple phase variable, which gives similar results as those shown in Fig.~\ref{F:Fdbk}D,E,F (data not shown). Similar approaches have been proposed previously~\cite{vincent2016driving}. These approaches can model the presence of time-keeping neural populations. For example, in songbird, motor learning is believed to be supported by a timekeeping signal from HVC, which is extensively used in models of songbird learning~\cite{doya1995novel,fiete2007model,fee2011hypothesis}. 

    \subsection*{Reward modulated learning with velocity control}
    In all examples considered so far, the output of the reservoir controlled the position of a pen or the angle of arm joints.  In control problems, motor output controls velocity or acceleration (e.g. applied force) of limbs or joints. From a naive perspective, SUPERTREX should still be able to complete such a task - random perturbations still change error, and SUPERTREX can learn to produce perturbations associated with lower error. 
    
    However, a more careful consideration reveals that SUPERTREX and RMHL applied directly to control velocity  would not be effective. To understand why, we first review and schematicize how SUPERTREX and RMHL successfully learn Task 1 where the output controls the \emph{position} of the pen, then consider why they would not work when the reservoir output control the \emph{velocity} of the pen. 
    
\begin{figure}
    \centering{
    \includegraphics[width = 4.5in]{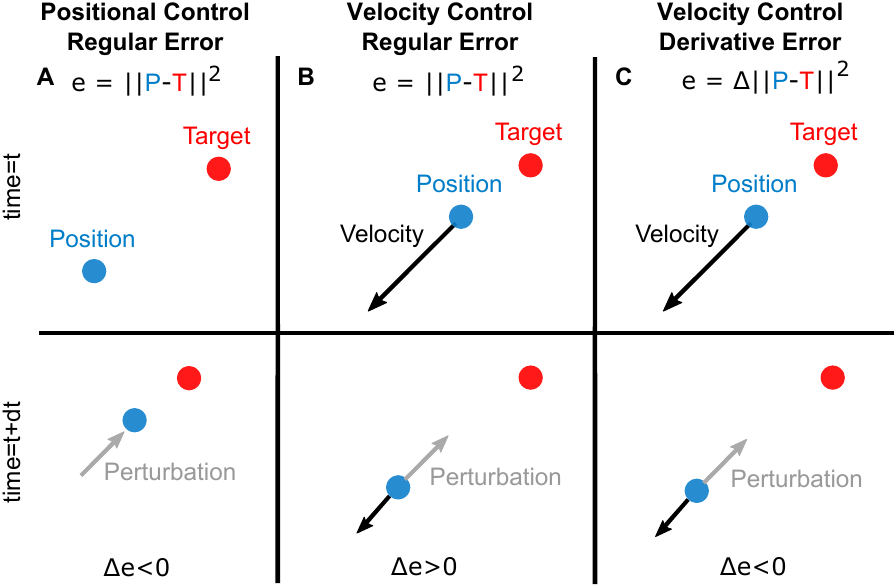}
    \caption{{\bf Velocity Control in SUPERTREX.} {\bf A)} When the position of a pen is controlled and error is the distance of the pen from its target, a beneficial perturbation correctly results in a decreased error.  {\bf B)} When the velocity of the pen is controlled and error is computed in the same way, there are situations where a beneficial perturbation results in increased error.
    {\bf C)} With velocity control, replacing the error by the derivative of the distance causes beneficial perturbations to correctly produce decreased error. 
    }
    \label{F:VelSchem}
    }
    \end{figure}

    In Task 1, suppose the pen is displaced from its target (Fig.~\ref{F:VelSchem} top left) and an exploratory perturbation is made to the reservoir output that successfully moves the pen closer to its target (Fig.~\ref{F:VelSchem} bottom left). In this case, the change in error is negative ($\Delta e\approx \hat e<0$), so the perturbation is correctly rewarded (see Eqs.~\eqref{E:dWRMHL} and \eqref{E:dWST}).

    Now consider Task 1 except that the reservoir output controls the velocity of the pen instead of the position. Again, suppose the pen is displaced from its target and also suppose that it is moving away from the target (Fig.~\ref{F:VelSchem} top middle). A beneficial exploratory perturbation changes the velocity of the pen in the direction of the target (Fig.~\ref{F:VelSchem} bottom middle). However, if the perturbation was not strong enough to change the direction of the pen, then the error (which is measured as the distance of the pen from its target) will still have increased after the perturbation (as in Fig.~\ref{F:VelSchem} bottom middle), so that $\Delta e\approx \hat e>0$ and this perturbation will be penalized instead of rewarded (as again indicated by Eqs.~\eqref{E:dWRMHL} and \eqref{E:dWST}). 
   
This problem is overcome by taking the derivative of the error, specifically defining $e$ to be the derivative of the distance between the pen and its target. When this change is made, then a reservoir controlling pen velocity will be correctly rewarded for beneficial perturbations Fig.~\ref{F:VelSchem} right) and penalized for harmful perturbations. 
    

            \begin{figure}
    \centering{
    \includegraphics[width = 4.5in]{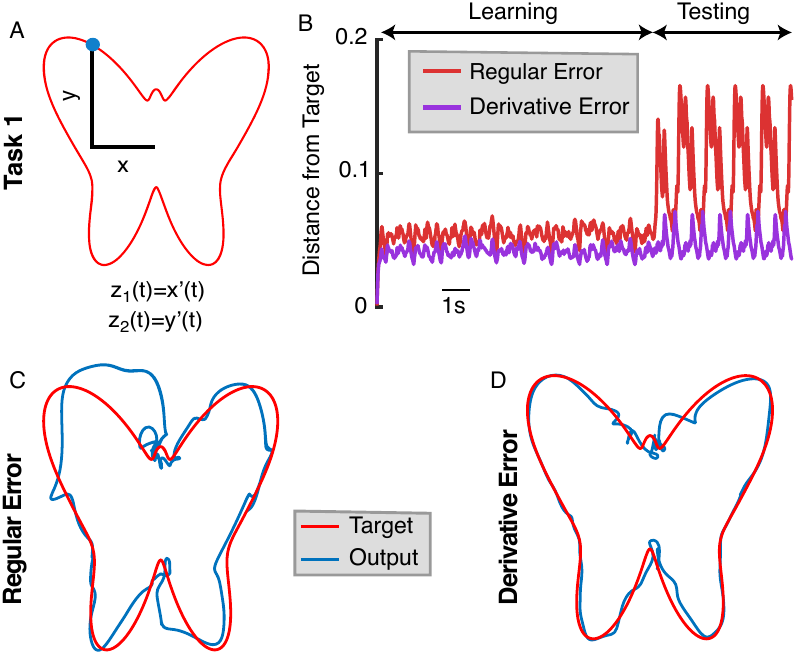}
    \caption{{\bf Velocity Control in SUPERTREX.} {\bf A)} A schematic of the velocity control task, which is identical to Task 1, except the velocity of the pen is controlled by the reservoir instead of its position.  {\bf B)} Using the regular error (distance of pen from target) produces large errors, but using the derivative of the distance produces smaller errors, especially during testing. {\bf C)} Butterfly drawn during the testing phase using regular error  and {\bf D)} derivative error.
    }
    \label{F:VelFig}
    }
    \end{figure}

       To test these conclusions, we repeated Task 1 with SUPERTREX, except with output now corresponding to velocity rather than position,
    \[
        \frac{d[x,y]}{dt} = \boldsymbol{z_1} + \boldsymbol{z_2}
    \]
    and we set $[x(0)\;y(0)] = \textbf{f}(0)$. During the course of training this model, we discovered two other adjustment were required. As our goal was to track a signal, rather than reach a target, adding a penalty term based on velocity was helpful in order to prevent oscillations around our target, e.g.
    \[
    \Perr=\Delta(\|{\bf f}-{\bf z}\|^2 + \gamma |dt \boldsymbol{z}|)
    \]
    Unfortunately, we did not find a systematic way to determine $\gamma$. Instead, $\gamma$ is chosen via iteration in order to prevent over- or under-damped oscillatory behavior. 
    
    Additionally, standard feedback $Q\textbf{z}$ clearly does not provide enough information - if we don't explicitly know our starting position, only knowing velocity does not help. Instead, we provided full-state information $Q[x\; y\; \textbf{f}]$ since we care more about our position than our velocity in terms of feedback. This is also more realistic - it makes sense to modify $\textbf{z}$ based on our position, rather than velocity, and position is more likely to be available as sensory feedback.
    Making these changes, we can compare SUPERTREX with error computed as the distance between the pen and its target (Fig.~\ref{F:VelFig}, ``regular error'') and with error computed as the derivative of the distance between the pen and its target (Fig.~\ref{F:VelFig}, ``derivative error''). As predicted, SUPERTREX with velocity control performs better when using the derivative of the distance as the error signal (Fig.~\ref{F:VelFig}, compare red to purple in panel B, and compare panel C to D).

    \section*{Discussion}
    

    
    We  presented a novel, reward-modulated method of reservoir computing, SUPERTREX, that performs nearly as well as fully supervised methods. This is desirable as there are a broad class or problems where traditional supervised methods are not applicable, such as Tasks 2 and 3 that we considered. Moreover, humans can learn motor tasks from reinforcement signals alone~\cite{Izawa2011}. 
    In place of a supervised error signal, SUPERTREX bootstraps from a dopamine-like, scalar error signal to a full error signal using rewarded exploration. This serves as an approximate target solution which is then transferred to a more traditional reservoir learning algorithm. This transfer of learned behavior to a mastery pathway, along with continued rewarded exploration, automatically creates a balanced system where the total output is correct, but the composition shifts over time from exploration to mastery. SUPERTREX performed similarly to FORCE on tasks where both were applicable, but also worked well on tasks where FORCE was not applicable. SUPERTREX also outperformed RMHL, a previously developed reward-modulated algorithm, on all tasks we considered.
    
    Unlike RMHL and other reinforcement learning models, SUPERTREX  models the complementary roles of cortical and basal ganglia pathways in motor learning. Under this interpretation, dopamine concentrations play the role of the reward signal, and the basal ganglia is the site of the RMHL-like, exploratory learning. Direct intra-cortical connections would then learn from Hebbian plasticity in the mastery pathway. Consistent with this interpretation, SUPERTREX produces inaccurate motor output when the reward signal is corrupted, modeling dopamine depletion in PD, but recovers the generation of well-learned output when the exploratory pathway is removed, modeling basal ganglia lesions used to treat PD. Hence SUPERTREX provides a model for understanding the role of motor learning in PD and its treatments.

    As models of motor learning, reward-modulated algorithms like SUPERTREX and RMHL assume no knowledge of the relationship between motor output and error. In contrast, fully supervised algorithms like FORCE require perfect  knowledge of this relationship. In reality, we learn through some combination of supervisory and reward-modulated error signals~\cite{Izawa2011}. 
    To account for this, SUPERTREX could potentially be extended to incorporate both one-dimensional reward and higher-dimensional sensory feedback.

    The FORCE-like learning algorithm used for the mastery pathway of SUPERTREX is biologistically unrealstic in some ways. The presence of the matrix, $P$, causes the rule to be non-local. However, we showed that SUPERTREX still works when $P$ is removed to implement a local, LMS learning rule (Fig.~\ref{F:PARK}). Indeed, one can replace the mastery pathway with any supervised learning rule. This could open the way for an implementation of SUPERTREX with spiking neural networks using existing supervised learning rules~\cite{Maass2002,bourdoukan2015enforcing,Abbott2016,pyle2017spatiotemporal}. In order to have a fully spiking-based version of SUPERTREX, this would also require a spike-based reinforcement learning rule, most likely an eligibility-trace based rule ~\cite{seung2003learning,xie2004learning,fiete2006gradient,miconi2017biologically}.
    
    As with most other reservoir computing algorithms, SUPERTREX implements online learning in which a local error signal is provided and used at every time step. This is partly by design - SUPERTREX learns extremely (even unrealistically) quickly as weights are updated at a high frequency. This learning is slowed by some extent by switching to the more realistic LMS learning rule (as in Fig.~\ref{F:PARK}). For some biological learning tasks, however, error signals are temporally sparse or reflect temporally non-local information.  Trial-based learning rules for reservoir computing~\cite{fiete2006gradient,miconi2017biologically} are applicable in the presence of sparse or non-local rewards. At least one of these algorithms learns very slowly, requiring thousands of trials~\cite{miconi2017biologically}, which may be an inevitable consequence of learning from sparse rewards. In reality, biological motor learning likely makes use of both online and sparse feedback. An extension of SUPERTREX that accounts for both types of feedback could be more versatile and realistic. 
    
    
    
    SUPERTREX is conceptually an extension of SPEED~\cite{Ashby2007}, which has a similar framework for categorization and other discrete tasks. SPEED  learns to map arbitrary discrete inputs to discrete outputs, such as in categorization tasks. While the architecture and learning rule are similar to SUPERTREX, SPEED cannot produce continuous, dynamical output and requires a separate pathway for each possible input-output pairing. 
    
    SUPERTREX could also be compared to a class of RNN algorithms that use a teacher network to train the final output network. However, many of these networks use the activity of the teacher network as a way to train the recurrence $J$ of the output network; in SUPERTREX, there is only one recurrent network (used for both outputs). These methods are often even more biologically implausible - for example, the recent FULL-FORCE extension of FORCE ~\cite{depasquale2018full} feeds the target signal info the first, chaotic reservoir, and then uses the activities of each reservoir unit in the teacher network as a target for training the second network, drastically increasing the amount of supervision required.
    
    SUPERTREX loses accuracy when learning is halted when feedback consists solely of the system's output (Fig.~\ref{F:Fdbk}A-C) due to the fact that it learns from a noisy estimate of the target. This shortcoming can be overcome by augmenting the feedback with the target, allowing the system to learn to self-correct noise-induced errors (Fig.~\ref{F:Fdbk}D-F).  FORCE is susceptible to the same instabilities as SUPERTREX under the biologically realistic assumption of noise during learning (Fig.~\ref{F:Fdbk}A), but SUPERTREX can solve tasks that FORCE cannot (Figs.~\ref{F:Task2} and \ref{F:Task3}). RMHL is also susceptible to the same instabilities and is applicable to the same tasks as SUPERTREX, but the instabilities in RMHL are not resolved by including target information in the feedback as they are for SUPERTREX (Fig.~\ref{F:Fdbk}C,D). Hence, SUPERTREX is the only one of the three algorithms that can be applied to reward-modulated learning tasks and achieves stability with target information in the feedback. Stability in reward-modulated reservoir computing without target information in the feedback term remains an open problem. This problem could potentially be solved by providing external input in-phase with the target output. This could help the reservoir ``keep time'' by re-aligning the reservoirs' state on each trial, allowing the system to self-correct its phase. A similar approach was shown to improve robustness of FORCE to perturbations in previous work~\cite{vincent2016driving}.
    
     Interestingly, biology may have already solved this problem. Research by Toledo-Suarez, Duarte, and Morrison~\cite{toledo2014liquid} has found that the striatum may act as a reservoir computer that processes state information. Rather than rely on raw inputs, the motor learning system instead has access to pre-processed state information that is both simpler and more relevant. In SUPERTREX, this could correspond to replacing our simple feedback of raw state information $Qz$ or $Q[z f]$ with $Qs$, where $s$ is a pre-processed state information vector. $s$ could even come from another reservoir, designed to ensure $s$ contains maximally relevant information to the task at hand. This would be an interesting extension to SUPERTREX.

    
    In summary, SUPERTREX is a new biologically inspired framework for reservoir computing that is more realistic and more effective than its predecessors. Using a general error signal allows for SUPERTREX to be used in places where a more powerful algorithm like FORCE cannot. The hand off from exploration to mastery allows SUPERTREX to perform nearly as well as FORCE with the generality of reward-modulated algorithms. Moreover, SUPERTREX offers a computational formalization of widely supported theories of motor learning and reproduces several experimental and clinical findings. Hence, this new framework opens the way for a truly two-way communication between biological and computational theories of motor learning.

    \section*{Materials and methods}

    \subsection*{Simulation and Reservoir Parameters}
    All simulations were performed using a forward Euler method, with $dt = 0.2$ms. Each task period or ``trial'' was $10^4$ ms long and all simulations except those in Fig.~\ref{F:PARK} had 15 trials. Fig.~\ref{F:PARK} had 110 trials. 
    
    The  reservoir equation used in all algorithms was
    \[
    \begin{aligned} 
    \tau \frac{d\textbf{x}}{dt}= -\textbf{x} + J\textbf{r}+ Q \textbf{z} 
    \end{aligned}
    \]
    where $\textbf{r}=\tanh(\textbf{x})+ \alpha\boldsymbol{\eta}$,  $\boldsymbol{\eta}$ was uniformly drawn from $[-1,1]$ on every time step, $\tau$ =10, and $\alpha = 2.5\times 10^{-2}$ during training and $\alpha=0$ during testing.  
    Reservoir size was set to $N = 1000$ neurons, with connection probability $p = 0.1$. Connection strengths in $J$ were normally distributed with mean 0 and variance ${\lambda^2}/({pN})$ with $\lambda = 1.5$. Feedback $Q$ was dense, with weights uniformly between $-1$ and $1$. Initial readout weights for RMHL and SUPERTREX exploratory pathway, as well as weights for FORCE and the SUPERTREX mastery pathway were initialized at $0$. Initial voltages were set uniformly between $-0.5$ and $0.5$, while initial rates were the hyperbolic tangent of initial voltages. Displayed outputs and errors were low pass filtered according to 
    \[
    \begin{aligned}
    \tau_{MSE}\frac{d\overline{MSE}(t)}{dt} &= -\overline{MSE}(t) + MSE(t) \\
    \tau_{bar} \frac{d\overline{\textbf{z}}(t)}{dt} &= -\overline{\textbf{z}}(t) + \textbf{z}(t)
    \end{aligned}
    \]
    where $\tau_{MSE} = 1000$, $\tau_{bar} = 10$, and $\overline{x}$ represents a low pass filtered version of the variable $x$. The plotted ``distance from target'' was computed as $\sqrt{\overline{MSE}}$ where $MSE(t)$ is the squared distance of the pen from its target. 
    
    \subsection*{FORCE}

    Reservoir output was 
    $
    {\bf z} = W {\bf r} 
    $ 
    and the learning rule is
    \[
    \tau_w\frac{dW}{dt} = - [{\bf z} - {\bf f}]{\bf r}^TP
    \]
    with $\tau_w$ = 0.02.
    The matrix $P$ is a running estimate of the inverse of the correlation matrix of rates $\bf r$, initialized to 
    \[
    P(0) = \frac{1}{\gamma}I
    \]
    and updated according to
    \[
    \tau_p\frac{dP}{dt} = - \frac{P\textbf{r}\,\textbf{r}^TP}{1 + \textbf{r}^TP\textbf{r}}
    \]
    where  $\tau_P = dt$, $\gamma=10$ is a constant and $I$ is the identity matrix. The matrix $P$ is only updated every 10 time steps in order to save on computing time.
    
    \subsection*{RMHL}
    For RMHL, outputs were given by 
    \[
    \begin{aligned}
    {\bf z} = W {\bf r} + \Psi(e)\boldsymbol{\eta}
    \end{aligned}
    \]
    and the learning rule was
    \[
    \tau_w\frac{dW}{dt} = \Phi(\hat{e})(\hat{{\bf z}}){\bf r}^T
    \]
    where $\tau_w=0.02$, $\eta$ is uniformly distributed noise between $[-1,1]$, and the high-pass filtered version, $\hat{x}$, of variable $x$ was computed as
    \[
    \begin{aligned}
    \tau\frac{d\overline{x}}{dt} &= -\overline{x} + x \\
    \hat{x} &= x - \overline{x}.
    \end{aligned}
    \]
    with $\tau = 1$ used for all tasks and trials. 
    
    \subsection*{SUPERTREX}
    Updates to $P$ were identical to the method used in FORCE above. Relevant other changes are
    \[
    \begin{aligned}
    {\bf z}_1 &= W_1 {\bf r} + \Psi(e)\boldsymbol{\eta} \\
    {\bf z}_2 &= W_2 {\bf r} \\
    {\bf z} &= {\bf z}_1 + {\bf z}_2
    \end{aligned}
    \]
    for $\boldsymbol{\eta}$ uniformly drawn from  $[-1, 1]$. For the learning algorithm,
    \[
    \begin{aligned}
    \tau_w\frac{dW_1}{dt} &= \Phi(\hat{e})\hat{\bf{z}}{\bf r}^T \\
    \tau_w\frac{dW_2}{dt} &= - k\bf{z_1}{\bf r}^TP
    \end{aligned}
    \]
    with $\tau_w$ remaining $0.02$ and constant learning rate $k$ which varies per task. Finally, an extra condition was imposed on updates to $P$, $W_2$. Both updates were multiplied by $(-0.5*\tanh(5\times10^5*(\overline{\Perr}-(1.5\times10^{-3})))+0.5)$, which acts as a soft threshold around $\Perr = 1.5\times10^{-3}$. Effectively, for errors larger than this the mastery pathway would not activate. Performance was similar, but slightly slower, without this thresholding.

    \subsection*{Tasks}
    
    In all tasks, the target was to draw a butterfly, given by a polar curve $x(t)=r(t)\cos(t)$ and $y(t)=r(t)\sin(t)$ where 
    \[
    \begin{aligned}
    r(t) = &c[9 - \sin(qt) + 2\sin(3qt) +\\ &2\sin(5qt) - \sin(7qt) + 3\cos(2qt) - 2\cos(4qt)]
    \end{aligned}
    \]
    and $c=1/\max_t[r(t)]$ is a normalizing constant. For a single repetition, t went from $0$ to $10^4$ms, and $q=\frac{2\pi}{10^4}$ scales the system such that $qt$ goes from $0$ to $2\pi$ over the duration.
    
    In task 2, the task is instead to draw a butterfly by controlling two angles, representing radians from y axis, and radians from the first joint. The arm is positioned at (0,-2), and each arm segment has fixed length of 1.8.  $h({\bf z})$ is therefore 
    \[
    \begin{aligned}
        h({\bf z}) = \left[\begin{matrix}
      1.8\sin(z_1 \pi) + 1.8\sin((z_1+z_2)\pi)  \\
      -2 + 1.8\cos(z_1\pi) + 1.8\cos((z_1+z_2)\pi)
     \end{matrix}  \right]
    \end{aligned}
    \]
    
    In task 3, now there are three angles to control. The arm is positioned at (0,-2), and each arm segment has fixed length. The first segment has length 1.8, the next 1.2, and the final .6. $h({\bf z})$ is therefore 
    \[
        h({\bf z}) = \left[\begin{matrix}
      1.8\sin(z_1 \pi) + 1.2\sin((z_1+z_2)\pi) + .6\sin((z_1+z_2+z_3)\pi) \\
      -2 + 1.8\cos(z_1\pi) + 1.2\cos((z_1+z_2)\pi) + .6\cos((z_1+z_2+z_3)\pi)
     \end{matrix} \right]
    \]
    
    In the first task, $\Psi(x) = 0.025\times\sqrt[4]{10x}$ and $\Phi(x) = -5\sqrt[4]{x}$ for both RMHL and SUPERTREX. When testing for swapping targets, $\Psi(x) = 0.1\times(-5x)^{0.3}$ and $\Phi(x) = 2.5\times\sqrt[4]{x}$. For SUPERTREX, learning rate $k$ was 0.5.

    For the second task, SUPERTREX learning rate $k$ was still 0.5, $\Psi(x) = 0.01\times\sqrt[5]{10x}$, and $\Phi(x) = 5\times\sqrt[4]{x}$.
    
    For the third task, SUPERTREX learning rate was $k=0.9$, $\Psi(x) = .025\times\sqrt[4]{10x}$, and $\Phi(x) = 5\times\sqrt[4]{x}$. The error metric was changed slightly, to
    \[
    e = ||h({\bf z}) - \textbf{f}||_2 + \alpha |\hat{{\bf z}_1}| + \beta |\hat{{\bf z}_2}| + \gamma |\hat{{\bf z}_3}|
    \]
    for $\alpha = 0.1$, $\beta = 0.05$, $\gamma = 0$. This implemented an additional cost for moving joints; highest for the longest arm segment, and 0 for the smallest arm segment.

    For the corrupted learning example, LMS learning was used, which is obtained by setting $P = I$. The learning rate was changed to $k=0.003$. Note that LMS learning rather than RMS learning generally requires a much lower learning rate. Other parameter values were the same as in the first task. The perturbation, $p(t)$, increased linearly from $0$ to $0.1$ over the corrupted learning timeframe.
    
    
    For the velocity controlled example in Fig.~\ref{F:VelSchem}, more significant changes were needed. As detailed, 
    \[
        [x,y](0) = f(0)
    \]
    and
    \[
        \frac{d[x,y]}{dt} = \boldsymbol{z_1} + \boldsymbol{z_2}
    \]
    as well as using full state feedback $Q[x\; y\; \textbf{f}]$. Learning rate $k$ was .025, as smaller velocities were needed relative to direct control of output. Velocity penalty $\gamma$ = $.3$. $\Psi$ and $\Phi$ were the same as in task 1. Finally, we changed how we we calculated $\hat{e}$. Rather than use a high pass filter as a crude derivative estimator, we instead used a finite difference approximation $\hat{e} = e(t) - e(t-dt)$. Note that, as described above, $e(t)$ now refers to $\Delta d(t) = d(t) - d(t-dt)$ where $d(t) = \|{\bf f}-{\bf z}\|^2 + \gamma |dt \boldsymbol{z}|$, e.g. the squared euclidean distance between the position and target plus a penalty term. Thus, our total update metric $\hat{e} = d(t) - 2d(t-dt) + d(t-2 dt)$, or the finite difference approximation to the second derivative of our error metric, which is euclidean distance plus penalty.
    
    \section*{Acknowledgments} 
    This work was supported by National Science Foundation grants DMS-1517828, DMS-1654268, and DBI-1707400. The authors thank Jonathan Rubin, Robert Turner, and Robert Mendez for helpful comments.

\bibliography{SupertrexBib}

\begin{thebibliography}{}

\bibitem [\protect \citeauthoryear {%
Abbott%
, Depasquale%
\BCBL {}\ \BBA {} Memmesheimer%
}{%
Abbott%
\ \protect \BOthers {.}}{%
{\protect \APACyear {2016}}%
}]{%
Abbott2016}
\APACinsertmetastar {%
Abbott2016}%
\begin{APACrefauthors}%
Abbott, L\BPBI F.%
, Depasquale, B.%
\BCBL {}\ \BBA {} Memmesheimer, R\BHBI m.%
\end{APACrefauthors}%
\unskip\
\newblock
\APACrefYearMonthDay{2016}{}{}.
\newblock
{\BBOQ}\APACrefatitle {{Building Functional Networks of Spiking Model Neurons}}
  {{Building Functional Networks of Spiking Model Neurons}}.{\BBCQ}
\newblock
\APACjournalVolNumPages{Nat. Neurosci.}{19}{3}{1--16}.
\PrintBackRefs{\CurrentBib}

\bibitem [\protect \citeauthoryear {%
Andalman%
\ \BBA {} Fee%
}{%
Andalman%
\ \BBA {} Fee%
}{%
{\protect \APACyear {2009}}%
}]{%
Andalman2009}
\APACinsertmetastar {%
Andalman2009}%
\begin{APACrefauthors}%
Andalman, A\BPBI S.%
\BCBT {}\ \BBA {} Fee, M\BPBI S.%
\end{APACrefauthors}%
\unskip\
\newblock
\APACrefYearMonthDay{2009}{}{}.
\newblock
{\BBOQ}\APACrefatitle {{A basal ganglia-forebrain circuit in the songbird
  biases motor output to avoid vocal errors.}} {{A basal ganglia-forebrain
  circuit in the songbird biases motor output to avoid vocal errors.}}{\BBCQ}
\newblock
\APACjournalVolNumPages{Proc. Natl. Acad. Sci. U. S. A.}{106}{30}{12518--23}.
\PrintBackRefs{\CurrentBib}

\bibitem [\protect \citeauthoryear {%
Aronov%
, Andalman%
\BCBL {}\ \BBA {} Fee%
}{%
Aronov%
\ \protect \BOthers {.}}{%
{\protect \APACyear {2008}}%
}]{%
Aronov2008}
\APACinsertmetastar {%
Aronov2008}%
\begin{APACrefauthors}%
Aronov, D.%
, Andalman, A.%
\BCBL {}\ \BBA {} Fee, M.%
\end{APACrefauthors}%
\unskip\
\newblock
\APACrefYearMonthDay{2008}{}{}.
\newblock
{\BBOQ}\APACrefatitle {{A specialized forebrain circuit for vocal babbling in
  the juvenile songbird}} {{A specialized forebrain circuit for vocal babbling
  in the juvenile songbird}}.{\BBCQ}
\newblock
\APACjournalVolNumPages{Science (80-. ).}{320}{}{630--635}.
\PrintBackRefs{\CurrentBib}

\bibitem [\protect \citeauthoryear {%
Ashby%
, Ennis%
\BCBL {}\ \BBA {} Spiering%
}{%
Ashby%
\ \protect \BOthers {.}}{%
{\protect \APACyear {2007}}%
}]{%
Ashby2007}
\APACinsertmetastar {%
Ashby2007}%
\begin{APACrefauthors}%
Ashby, F\BPBI G.%
, Ennis, J\BPBI M.%
\BCBL {}\ \BBA {} Spiering, B\BPBI J.%
\end{APACrefauthors}%
\unskip\
\newblock
\APACrefYearMonthDay{2007}{}{}.
\newblock
{\BBOQ}\APACrefatitle {{A neurobiological theory of automaticity in perceptual
  categorization.}} {{A neurobiological theory of automaticity in perceptual
  categorization.}}{\BBCQ}
\newblock
\APACjournalVolNumPages{Psychol. Rev.}{114}{3}{632--56}.
\PrintBackRefs{\CurrentBib}

\bibitem [\protect \citeauthoryear {%
Ashby%
, Turner%
\BCBL {}\ \BBA {} Horvitz%
}{%
Ashby%
\ \protect \BOthers {.}}{%
{\protect \APACyear {2010}}%
}]{%
ashby2010cortical}
\APACinsertmetastar {%
ashby2010cortical}%
\begin{APACrefauthors}%
Ashby, F\BPBI G.%
, Turner, B\BPBI O.%
\BCBL {}\ \BBA {} Horvitz, J\BPBI C.%
\end{APACrefauthors}%
\unskip\
\newblock
\APACrefYearMonthDay{2010}{}{}.
\newblock
{\BBOQ}\APACrefatitle {Cortical and basal ganglia contributions to habit
  learning and automaticity} {Cortical and basal ganglia contributions to habit
  learning and automaticity}.{\BBCQ}
\newblock
\APACjournalVolNumPages{Trends Cog. Sci.}{14}{5}{208--215}.
\PrintBackRefs{\CurrentBib}

\bibitem [\protect \citeauthoryear {%
Bottjer%
, Miesner%
\BCBL {}\ \BBA {} Arnold%
}{%
Bottjer%
\ \protect \BOthers {.}}{%
{\protect \APACyear {1984}}%
}]{%
Bottjer1984}
\APACinsertmetastar {%
Bottjer1984}%
\begin{APACrefauthors}%
Bottjer, S\BPBI W.%
, Miesner, E\BPBI a.%
\BCBL {}\ \BBA {} Arnold, a\BPBI P.%
\end{APACrefauthors}%
\unskip\
\newblock
\APACrefYearMonthDay{1984}{}{}.
\newblock
{\BBOQ}\APACrefatitle {{Forebrain lesions disrupt development but not
  maintenance of song in passerine birds.}} {{Forebrain lesions disrupt
  development but not maintenance of song in passerine birds.}}{\BBCQ}
\newblock
\APACjournalVolNumPages{Science (80-. ).}{224}{4651}{901--903}.
\PrintBackRefs{\CurrentBib}

\bibitem [\protect \citeauthoryear {%
Bourdoukan%
\ \BBA {} Deneve%
}{%
Bourdoukan%
\ \BBA {} Deneve%
}{%
{\protect \APACyear {2015}}%
}]{%
bourdoukan2015enforcing}
\APACinsertmetastar {%
bourdoukan2015enforcing}%
\begin{APACrefauthors}%
Bourdoukan, R.%
\BCBT {}\ \BBA {} Deneve, S.%
\end{APACrefauthors}%
\unskip\
\newblock
\APACrefYearMonthDay{2015}{}{}.
\newblock
{\BBOQ}\APACrefatitle {Enforcing balance allows local supervised learning in
  spiking recurrent networks} {Enforcing balance allows local supervised
  learning in spiking recurrent networks}.{\BBCQ}
\newblock
\BIn{} \APACrefbtitle {Adv. Neur. In.} {Adv. neur. in.}\ (\BPGS\ 982--990).
\PrintBackRefs{\CurrentBib}

\bibitem [\protect \citeauthoryear {%
Brainard%
\ \BBA {} Doupe%
}{%
Brainard%
\ \BBA {} Doupe%
}{%
{\protect \APACyear {2000}}%
}]{%
Brainard2000}
\APACinsertmetastar {%
Brainard2000}%
\begin{APACrefauthors}%
Brainard%
\BCBT {}\ \BBA {} Doupe, A.%
\end{APACrefauthors}%
\unskip\
\newblock
\APACrefYearMonthDay{2000}{}{}.
\newblock
{\BBOQ}\APACrefatitle {{Interruption of a basal ganglia–forebrain circuit
  prevents plasticity of learned vocalizations}} {{Interruption of a basal
  ganglia–forebrain circuit prevents plasticity of learned
  vocalizations}}.{\BBCQ}
\newblock
\APACjournalVolNumPages{Nature}{404}{}{}.
\PrintBackRefs{\CurrentBib}

\bibitem [\protect \citeauthoryear {%
Brainard%
\ \BBA {} Doupe%
}{%
Brainard%
\ \BBA {} Doupe%
}{%
{\protect \APACyear {2002}}%
}]{%
Brainard2002}
\APACinsertmetastar {%
Brainard2002}%
\begin{APACrefauthors}%
Brainard%
\BCBT {}\ \BBA {} Doupe, A.%
\end{APACrefauthors}%
\unskip\
\newblock
\APACrefYearMonthDay{2002}{}{}.
\newblock
{\BBOQ}\APACrefatitle {{What songbirds teach us about learning}} {{What
  songbirds teach us about learning}}.{\BBCQ}
\newblock
\APACjournalVolNumPages{Nature}{417}{}{351--358}.
\PrintBackRefs{\CurrentBib}

\bibitem [\protect \citeauthoryear {%
M\BPBI S.~Brainard%
}{%
M\BPBI S.~Brainard%
}{%
{\protect \APACyear {2004}}%
}]{%
brainard2004contributions}
\APACinsertmetastar {%
brainard2004contributions}%
\begin{APACrefauthors}%
Brainard, M\BPBI S.%
\end{APACrefauthors}%
\unskip\
\newblock
\APACrefYearMonthDay{2004}{}{}.
\newblock
{\BBOQ}\APACrefatitle {Contributions of the anterior forebrain pathway to vocal
  plasticity} {Contributions of the anterior forebrain pathway to vocal
  plasticity}.{\BBCQ}
\newblock
\APACjournalVolNumPages{Ann. NY Acad. Sci.}{1016}{1}{377--394}.
\PrintBackRefs{\CurrentBib}

\bibitem [\protect \citeauthoryear {%
Carelli%
, Wolske%
\BCBL {}\ \BBA {} West%
}{%
Carelli%
\ \protect \BOthers {.}}{%
{\protect \APACyear {1997}}%
}]{%
Carelli1997}
\APACinsertmetastar {%
Carelli1997}%
\begin{APACrefauthors}%
Carelli, R\BPBI M.%
, Wolske, M.%
\BCBL {}\ \BBA {} West, M\BPBI O.%
\end{APACrefauthors}%
\unskip\
\newblock
\APACrefYearMonthDay{1997}{}{}.
\newblock
{\BBOQ}\APACrefatitle {{Loss of lever press-related firing of rat striatal
  forelimb neurons after repeated sessions in a lever pressing task}} {{Loss of
  lever press-related firing of rat striatal forelimb neurons after repeated
  sessions in a lever pressing task}}.{\BBCQ}
\newblock
\APACjournalVolNumPages{J. Neurosci.}{17}{5}{1804--1814}.
\PrintBackRefs{\CurrentBib}

\bibitem [\protect \citeauthoryear {%
Churchland%
\ \protect \BOthers {.}}{%
Churchland%
\ \protect \BOthers {.}}{%
{\protect \APACyear {2012}}%
}]{%
Churchland2012}
\APACinsertmetastar {%
Churchland2012}%
\begin{APACrefauthors}%
Churchland, M\BPBI M.%
, Cunningham, J\BPBI P.%
, Kaufman, M\BPBI T.%
, Foster, J\BPBI D.%
, Nuyujukian, P.%
, Ryu, S\BPBI I.%
\BCBL {}\ \BBA {} Shenoy, K\BPBI V.%
\end{APACrefauthors}%
\unskip\
\newblock
\APACrefYearMonthDay{2012}{}{}.
\newblock
{\BBOQ}\APACrefatitle {{Neural population dynamics during reaching.}} {{Neural
  population dynamics during reaching.}}{\BBCQ}
\newblock
\APACjournalVolNumPages{Nature}{487}{7405}{51--6}.
\PrintBackRefs{\CurrentBib}

\bibitem [\protect \citeauthoryear {%
DePasquale%
, Cueva%
, Rajan%
, Abbott%
\BCBL {}\ \protect \BOthers {.}}{%
DePasquale%
\ \protect \BOthers {.}}{%
{\protect \APACyear {2018}}%
}]{%
depasquale2018full}
\APACinsertmetastar {%
depasquale2018full}%
\begin{APACrefauthors}%
DePasquale, B.%
, Cueva, C\BPBI J.%
, Rajan, K.%
, Abbott, L.%
\BCBL {}\ \BOthersPeriod {.}\end{APACrefauthors}%
\unskip\
\newblock
\APACrefYearMonthDay{2018}{}{}.
\newblock
{\BBOQ}\APACrefatitle {full-FORCE: A target-based method for training recurrent
  networks} {full-force: A target-based method for training recurrent
  networks}.{\BBCQ}
\newblock
\APACjournalVolNumPages{PloS one}{13}{2}{e0191527}.
\PrintBackRefs{\CurrentBib}

\bibitem [\protect \citeauthoryear {%
Doya%
\ \BBA {} Sejnowski%
}{%
Doya%
\ \BBA {} Sejnowski%
}{%
{\protect \APACyear {1995}}%
}]{%
doya1995novel}
\APACinsertmetastar {%
doya1995novel}%
\begin{APACrefauthors}%
Doya, K.%
\BCBT {}\ \BBA {} Sejnowski, T\BPBI J.%
\end{APACrefauthors}%
\unskip\
\newblock
\APACrefYearMonthDay{1995}{}{}.
\newblock
{\BBOQ}\APACrefatitle {A novel reinforcement model of birdsong vocalization
  learning} {A novel reinforcement model of birdsong vocalization
  learning}.{\BBCQ}
\newblock
\BIn{} \APACrefbtitle {Adv. Neur. In.} {Adv. neur. in.}\ (\BPGS\ 101--108).
\PrintBackRefs{\CurrentBib}

\bibitem [\protect \citeauthoryear {%
Fee%
}{%
Fee%
}{%
{\protect \APACyear {2014}}%
}]{%
Fee2014}
\APACinsertmetastar {%
Fee2014}%
\begin{APACrefauthors}%
Fee, M\BPBI S.%
\end{APACrefauthors}%
\unskip\
\newblock
\APACrefYearMonthDay{2014}{}{}.
\newblock
{\BBOQ}\APACrefatitle {{The role of efference copy in striatal learning.}}
  {{The role of efference copy in striatal learning.}}{\BBCQ}
\newblock
\APACjournalVolNumPages{Curr. Opin. Neurobiol.}{25}{}{194--200}.
\PrintBackRefs{\CurrentBib}

\bibitem [\protect \citeauthoryear {%
Fee%
\ \BBA {} Goldberg%
}{%
Fee%
\ \BBA {} Goldberg%
}{%
{\protect \APACyear {2011}}%
}]{%
fee2011hypothesis}
\APACinsertmetastar {%
fee2011hypothesis}%
\begin{APACrefauthors}%
Fee, M\BPBI S.%
\BCBT {}\ \BBA {} Goldberg, J\BPBI H.%
\end{APACrefauthors}%
\unskip\
\newblock
\APACrefYearMonthDay{2011}{}{}.
\newblock
{\BBOQ}\APACrefatitle {A hypothesis for basal ganglia-dependent reinforcement
  learning in the songbird} {A hypothesis for basal ganglia-dependent
  reinforcement learning in the songbird}.{\BBCQ}
\newblock
\APACjournalVolNumPages{Neuroscience}{198}{}{152--170}.
\PrintBackRefs{\CurrentBib}

\bibitem [\protect \citeauthoryear {%
Fiete%
, Fee%
\BCBL {}\ \BBA {} Seung%
}{%
Fiete%
\ \protect \BOthers {.}}{%
{\protect \APACyear {2007}}%
}]{%
fiete2007model}
\APACinsertmetastar {%
fiete2007model}%
\begin{APACrefauthors}%
Fiete, I\BPBI R.%
, Fee, M\BPBI S.%
\BCBL {}\ \BBA {} Seung, H\BPBI S.%
\end{APACrefauthors}%
\unskip\
\newblock
\APACrefYearMonthDay{2007}{}{}.
\newblock
{\BBOQ}\APACrefatitle {Model of birdsong learning based on gradient estimation
  by dynamic perturbation of neural conductances} {Model of birdsong learning
  based on gradient estimation by dynamic perturbation of neural
  conductances}.{\BBCQ}
\newblock
\APACjournalVolNumPages{J. Neuropysiol.}{98}{4}{2038--2057}.
\PrintBackRefs{\CurrentBib}

\bibitem [\protect \citeauthoryear {%
Fiete%
\ \BBA {} Seung%
}{%
Fiete%
\ \BBA {} Seung%
}{%
{\protect \APACyear {2006}}%
}]{%
fiete2006gradient}
\APACinsertmetastar {%
fiete2006gradient}%
\begin{APACrefauthors}%
Fiete, I\BPBI R.%
\BCBT {}\ \BBA {} Seung, H\BPBI S.%
\end{APACrefauthors}%
\unskip\
\newblock
\APACrefYearMonthDay{2006}{}{}.
\newblock
{\BBOQ}\APACrefatitle {Gradient learning in spiking neural networks by dynamic
  perturbation of conductances} {Gradient learning in spiking neural networks
  by dynamic perturbation of conductances}.{\BBCQ}
\newblock
\APACjournalVolNumPages{Phys. Rev. Lett.}{97}{4}{048104}.
\PrintBackRefs{\CurrentBib}

\bibitem [\protect \citeauthoryear {%
H{\'{e}}lie%
, Paul%
\BCBL {}\ \BBA {} Ashby%
}{%
H{\'{e}}lie%
\ \protect \BOthers {.}}{%
{\protect \APACyear {2012}}%
}]{%
Helie2012}
\APACinsertmetastar {%
Helie2012}%
\begin{APACrefauthors}%
H{\'{e}}lie, S.%
, Paul, E\BPBI J.%
\BCBL {}\ \BBA {} Ashby, F\BPBI G.%
\end{APACrefauthors}%
\unskip\
\newblock
\APACrefYearMonthDay{2012}{}{}.
\newblock
{\BBOQ}\APACrefatitle {{A neurocomputational account of cognitive deficits in
  Parkinson's disease.}} {{A neurocomputational account of cognitive deficits
  in Parkinson's disease.}}{\BBCQ}
\newblock
\APACjournalVolNumPages{Neuropsychologia}{50}{9}{2290--302}.
\PrintBackRefs{\CurrentBib}

\bibitem [\protect \citeauthoryear {%
Hennequin%
, Vogels%
\BCBL {}\ \BBA {} Gerstner%
}{%
Hennequin%
\ \protect \BOthers {.}}{%
{\protect \APACyear {2014}}%
}]{%
Hennequin2014}
\APACinsertmetastar {%
Hennequin2014}%
\begin{APACrefauthors}%
Hennequin, G.%
, Vogels, T\BPBI P.%
\BCBL {}\ \BBA {} Gerstner, W.%
\end{APACrefauthors}%
\unskip\
\newblock
\APACrefYearMonthDay{2014}{}{}.
\newblock
{\BBOQ}\APACrefatitle {{Optimal control of transient dynamics in balanced
  networks supports generation of complex movements.}} {{Optimal control of
  transient dynamics in balanced networks supports generation of complex
  movements.}}{\BBCQ}
\newblock
\APACjournalVolNumPages{Neuron}{82}{6}{1394--406}.
\PrintBackRefs{\CurrentBib}

\bibitem [\protect \citeauthoryear {%
Hoerzer%
, Legenstein%
\BCBL {}\ \BBA {} Maass%
}{%
Hoerzer%
\ \protect \BOthers {.}}{%
{\protect \APACyear {2014}}%
}]{%
hoerzer2012emergence}
\APACinsertmetastar {%
hoerzer2012emergence}%
\begin{APACrefauthors}%
Hoerzer, G\BPBI M.%
, Legenstein, R.%
\BCBL {}\ \BBA {} Maass, W.%
\end{APACrefauthors}%
\unskip\
\newblock
\APACrefYearMonthDay{2014}{}{}.
\newblock
{\BBOQ}\APACrefatitle {Emergence of complex computational structures from
  chaotic neural networks through reward-modulated Hebbian learning} {Emergence
  of complex computational structures from chaotic neural networks through
  reward-modulated hebbian learning}.{\BBCQ}
\newblock
\APACjournalVolNumPages{Cereb. Cort.}{24}{3}{677--690}.
\PrintBackRefs{\CurrentBib}

\bibitem [\protect \citeauthoryear {%
Izawa%
\ \BBA {} Shadmehr%
}{%
Izawa%
\ \BBA {} Shadmehr%
}{%
{\protect \APACyear {2011}}%
}]{%
Izawa2011}
\APACinsertmetastar {%
Izawa2011}%
\begin{APACrefauthors}%
Izawa, J.%
\BCBT {}\ \BBA {} Shadmehr, R.%
\end{APACrefauthors}%
\unskip\
\newblock
\APACrefYearMonthDay{2011}{}{}.
\newblock
{\BBOQ}\APACrefatitle {{Learning from sensory and reward prediction errors
  during motor adaptation}} {{Learning from sensory and reward prediction
  errors during motor adaptation}}.{\BBCQ}
\newblock
\APACjournalVolNumPages{PLoS Comput. Biol.}{7}{3}{1--11}.
\PrintBackRefs{\CurrentBib}

\bibitem [\protect \citeauthoryear {%
Jaeger%
\ \BBA {} Haas%
}{%
Jaeger%
\ \BBA {} Haas%
}{%
{\protect \APACyear {2004}}%
}]{%
Jaeger2004}
\APACinsertmetastar {%
Jaeger2004}%
\begin{APACrefauthors}%
Jaeger, H.%
\BCBT {}\ \BBA {} Haas, H.%
\end{APACrefauthors}%
\unskip\
\newblock
\APACrefYearMonthDay{2004}{}{}.
\newblock
{\BBOQ}\APACrefatitle {{Harnessing nonlinearity: predicting chaotic systems and
  saving energy in wireless communication.}} {{Harnessing nonlinearity:
  predicting chaotic systems and saving energy in wireless
  communication.}}{\BBCQ}
\newblock
\APACjournalVolNumPages{Science}{304}{5667}{78--80}.
\PrintBackRefs{\CurrentBib}

\bibitem [\protect \citeauthoryear {%
Kao%
, Doupe%
\BCBL {}\ \BBA {} Brainard%
}{%
Kao%
\ \protect \BOthers {.}}{%
{\protect \APACyear {2005}}%
}]{%
Kao2005}
\APACinsertmetastar {%
Kao2005}%
\begin{APACrefauthors}%
Kao, M\BPBI H.%
, Doupe, A\BPBI J.%
\BCBL {}\ \BBA {} Brainard, M\BPBI S.%
\end{APACrefauthors}%
\unskip\
\newblock
\APACrefYearMonthDay{2005}{}{}.
\newblock
{\BBOQ}\APACrefatitle {Contributions of an avian basal ganglia--forebrain
  circuit to real-time modulation of song} {Contributions of an avian basal
  ganglia--forebrain circuit to real-time modulation of song}.{\BBCQ}
\newblock
\APACjournalVolNumPages{Nature}{433}{7026}{638--643}.
\PrintBackRefs{\CurrentBib}

\bibitem [\protect \citeauthoryear {%
Kawai%
\ \protect \BOthers {.}}{%
Kawai%
\ \protect \BOthers {.}}{%
{\protect \APACyear {2015}}%
}]{%
kawai2015motor}
\APACinsertmetastar {%
kawai2015motor}%
\begin{APACrefauthors}%
Kawai, R.%
, Markman, T.%
, Poddar, R.%
, Ko, R.%
, Fantana, A\BPBI L.%
, Dhawale, A\BPBI K.%
\BDBL {}{\"O}lveczky, B\BPBI P.%
\end{APACrefauthors}%
\unskip\
\newblock
\APACrefYearMonthDay{2015}{}{}.
\newblock
{\BBOQ}\APACrefatitle {Motor cortex is required for learning but not for
  executing a motor skill} {Motor cortex is required for learning but not for
  executing a motor skill}.{\BBCQ}
\newblock
\APACjournalVolNumPages{Neuron}{86}{3}{800--812}.
\PrintBackRefs{\CurrentBib}

\bibitem [\protect \citeauthoryear {%
Laje%
\ \BBA {} Buonomano%
}{%
Laje%
\ \BBA {} Buonomano%
}{%
{\protect \APACyear {2013}}%
}]{%
Laje2013}
\APACinsertmetastar {%
Laje2013}%
\begin{APACrefauthors}%
Laje, R.%
\BCBT {}\ \BBA {} Buonomano, D\BPBI V.%
\end{APACrefauthors}%
\unskip\
\newblock
\APACrefYearMonthDay{2013}{}{}.
\newblock
{\BBOQ}\APACrefatitle {{Robust timing and motor patterns by taming chaos in
  recurrent neural networks.}} {{Robust timing and motor patterns by taming
  chaos in recurrent neural networks.}}{\BBCQ}
\newblock
\APACjournalVolNumPages{Nat. Neurosci.}{16}{7}{925--33}.
\PrintBackRefs{\CurrentBib}

\bibitem [\protect \citeauthoryear {%
Luko{\v{s}}evi{\v{c}}ius%
, Jaeger%
\BCBL {}\ \BBA {} Schrauwen%
}{%
Luko{\v{s}}evi{\v{c}}ius%
\ \protect \BOthers {.}}{%
{\protect \APACyear {2012}}%
}]{%
Lukosevicius2012}
\APACinsertmetastar {%
Lukosevicius2012}%
\begin{APACrefauthors}%
Luko{\v{s}}evi{\v{c}}ius, M.%
, Jaeger, H.%
\BCBL {}\ \BBA {} Schrauwen, B.%
\end{APACrefauthors}%
\unskip\
\newblock
\APACrefYearMonthDay{2012}{}{}.
\newblock
{\BBOQ}\APACrefatitle {{Reservoir Computing Trends}} {{Reservoir Computing
  Trends}}.{\BBCQ}
\newblock
\APACjournalVolNumPages{KI - K{\"{u}}nstliche Intelligenz}{26}{4}{365--371}.
\PrintBackRefs{\CurrentBib}

\bibitem [\protect \citeauthoryear {%
Maass%
, Natschl{\"{a}}ger%
\BCBL {}\ \BBA {} Markram%
}{%
Maass%
\ \protect \BOthers {.}}{%
{\protect \APACyear {2002}}%
}]{%
Maass2002}
\APACinsertmetastar {%
Maass2002}%
\begin{APACrefauthors}%
Maass, W.%
, Natschl{\"{a}}ger, T.%
\BCBL {}\ \BBA {} Markram, H.%
\end{APACrefauthors}%
\unskip\
\newblock
\APACrefYearMonthDay{2002}{}{}.
\newblock
{\BBOQ}\APACrefatitle {{Real-time computing without stable states: a new
  framework for neural computation based on perturbations.}} {{Real-time
  computing without stable states: a new framework for neural computation based
  on perturbations.}}{\BBCQ}
\newblock
\APACjournalVolNumPages{Neural Comput.}{14}{11}{2531--60}.
\PrintBackRefs{\CurrentBib}

\bibitem [\protect \citeauthoryear {%
Mante%
, Sussillo%
, Shenoy%
\BCBL {}\ \BBA {} Newsome%
}{%
Mante%
\ \protect \BOthers {.}}{%
{\protect \APACyear {2013}}%
}]{%
Mante2013}
\APACinsertmetastar {%
Mante2013}%
\begin{APACrefauthors}%
Mante, V.%
, Sussillo, D.%
, Shenoy, K\BPBI V.%
\BCBL {}\ \BBA {} Newsome, W\BPBI T.%
\end{APACrefauthors}%
\unskip\
\newblock
\APACrefYearMonthDay{2013}{}{}.
\newblock
{\BBOQ}\APACrefatitle {{Context-dependent computation by recurrent dynamics in
  prefrontal cortex.}} {{Context-dependent computation by recurrent dynamics in
  prefrontal cortex.}}{\BBCQ}
\newblock
\APACjournalVolNumPages{Nature}{503}{7474}{78--84}.
\PrintBackRefs{\CurrentBib}

\bibitem [\protect \citeauthoryear {%
Miconi%
}{%
Miconi%
}{%
{\protect \APACyear {2017}}%
}]{%
miconi2017biologically}
\APACinsertmetastar {%
miconi2017biologically}%
\begin{APACrefauthors}%
Miconi, T.%
\end{APACrefauthors}%
\unskip\
\newblock
\APACrefYearMonthDay{2017}{}{}.
\newblock
{\BBOQ}\APACrefatitle {Biologically plausible learning in recurrent neural
  networks reproduces neural dynamics observed during cognitive tasks}
  {Biologically plausible learning in recurrent neural networks reproduces
  neural dynamics observed during cognitive tasks}.{\BBCQ}
\newblock
\APACjournalVolNumPages{E-Life}{6}{}{}.
\PrintBackRefs{\CurrentBib}

\bibitem [\protect \citeauthoryear {%
Miyachi%
, Hikosaka%
\BCBL {}\ \BBA {} Lu%
}{%
Miyachi%
\ \protect \BOthers {.}}{%
{\protect \APACyear {2002}}%
}]{%
miyachi2002differential}
\APACinsertmetastar {%
miyachi2002differential}%
\begin{APACrefauthors}%
Miyachi, S.%
, Hikosaka, O.%
\BCBL {}\ \BBA {} Lu, X.%
\end{APACrefauthors}%
\unskip\
\newblock
\APACrefYearMonthDay{2002}{}{}.
\newblock
{\BBOQ}\APACrefatitle {Differential activation of monkey striatal neurons in
  the early and late stages of procedural learning} {Differential activation of
  monkey striatal neurons in the early and late stages of procedural
  learning}.{\BBCQ}
\newblock
\APACjournalVolNumPages{Exp Brain Res}{146}{1}{122--126}.
\PrintBackRefs{\CurrentBib}

\bibitem [\protect \citeauthoryear {%
Miyachi%
, Hikosaka%
, Miyashita%
, K{\'a}r{\'a}di%
\BCBL {}\ \BBA {} Rand%
}{%
Miyachi%
\ \protect \BOthers {.}}{%
{\protect \APACyear {1997}}%
}]{%
miyachi1997differential}
\APACinsertmetastar {%
miyachi1997differential}%
\begin{APACrefauthors}%
Miyachi, S.%
, Hikosaka, O.%
, Miyashita, K.%
, K{\'a}r{\'a}di, Z.%
\BCBL {}\ \BBA {} Rand, M\BPBI K.%
\end{APACrefauthors}%
\unskip\
\newblock
\APACrefYearMonthDay{1997}{}{}.
\newblock
{\BBOQ}\APACrefatitle {Differential roles of monkey striatum in learning of
  sequential hand movement} {Differential roles of monkey striatum in learning
  of sequential hand movement}.{\BBCQ}
\newblock
\APACjournalVolNumPages{Exp. brain res.}{115}{1}{1--5}.
\PrintBackRefs{\CurrentBib}

\bibitem [\protect \citeauthoryear {%
Obeso%
\ \protect \BOthers {.}}{%
Obeso%
\ \protect \BOthers {.}}{%
{\protect \APACyear {2009}}%
}]{%
Obeso2009}
\APACinsertmetastar {%
Obeso2009}%
\begin{APACrefauthors}%
Obeso, J\BPBI a.%
, Jahanshahi, M.%
, Alvarez, L.%
, Macias, R.%
, Pedroso, I.%
, Wilkinson, L.%
\BDBL {}Rothwell, J\BPBI C.%
\end{APACrefauthors}%
\unskip\
\newblock
\APACrefYearMonthDay{2009}{}{}.
\newblock
{\BBOQ}\APACrefatitle {{What can man do without basal ganglia motor output? The
  effect of combined unilateral subthalamotomy and pallidotomy in a patient
  with Parkinson's disease.}} {{What can man do without basal ganglia motor
  output? The effect of combined unilateral subthalamotomy and pallidotomy in a
  patient with Parkinson's disease.}}{\BBCQ}
\newblock
\APACjournalVolNumPages{Exp. Neurol.}{220}{2}{283--92}.
\PrintBackRefs{\CurrentBib}

\bibitem [\protect \citeauthoryear {%
Olveczky%
, Andalman%
\BCBL {}\ \BBA {} Fee%
}{%
Olveczky%
\ \protect \BOthers {.}}{%
{\protect \APACyear {2005}}%
}]{%
Olveczky2005}
\APACinsertmetastar {%
Olveczky2005}%
\begin{APACrefauthors}%
Olveczky, B\BPBI P.%
, Andalman, A\BPBI S.%
\BCBL {}\ \BBA {} Fee, M\BPBI S.%
\end{APACrefauthors}%
\unskip\
\newblock
\APACrefYearMonthDay{2005}{}{}.
\newblock
{\BBOQ}\APACrefatitle {{Vocal experimentation in the juvenile songbird requires
  a basal ganglia circuit.}} {{Vocal experimentation in the juvenile songbird
  requires a basal ganglia circuit.}}{\BBCQ}
\newblock
\APACjournalVolNumPages{PLoS Biol.}{3}{5}{e153}.
\PrintBackRefs{\CurrentBib}

\bibitem [\protect \citeauthoryear {%
{\"{O}}lveczky%
, Otchy%
, Goldberg%
, Aronov%
\BCBL {}\ \BBA {} Fee%
}{%
{\"{O}}lveczky%
\ \protect \BOthers {.}}{%
{\protect \APACyear {2011}}%
}]{%
Olveczky2011}
\APACinsertmetastar {%
Olveczky2011}%
\begin{APACrefauthors}%
{\"{O}}lveczky, B\BPBI P.%
, Otchy, T\BPBI M.%
, Goldberg, J\BPBI H.%
, Aronov, D.%
\BCBL {}\ \BBA {} Fee, M\BPBI S.%
\end{APACrefauthors}%
\unskip\
\newblock
\APACrefYearMonthDay{2011}{}{}.
\newblock
{\BBOQ}\APACrefatitle {{Changes in the neural control of a complex motor
  sequence during learning.}} {{Changes in the neural control of a complex
  motor sequence during learning.}}{\BBCQ}
\newblock
\APACjournalVolNumPages{J. Neurophysiol.}{106}{1}{386--97}.
\PrintBackRefs{\CurrentBib}

\bibitem [\protect \citeauthoryear {%
Pasupathy%
\ \BBA {} Miller%
}{%
Pasupathy%
\ \BBA {} Miller%
}{%
{\protect \APACyear {2005}}%
}]{%
Pasupathy2005}
\APACinsertmetastar {%
Pasupathy2005}%
\begin{APACrefauthors}%
Pasupathy, A.%
\BCBT {}\ \BBA {} Miller, E\BPBI K.%
\end{APACrefauthors}%
\unskip\
\newblock
\APACrefYearMonthDay{2005}{}{}.
\newblock
{\BBOQ}\APACrefatitle {Different time courses of learning-related activity in
  the prefrontal cortex and striatum} {Different time courses of
  learning-related activity in the prefrontal cortex and striatum}.{\BBCQ}
\newblock
\APACjournalVolNumPages{Nature}{433}{7028}{873--876}.
\PrintBackRefs{\CurrentBib}

\bibitem [\protect \citeauthoryear {%
Poldrack%
\ \protect \BOthers {.}}{%
Poldrack%
\ \protect \BOthers {.}}{%
{\protect \APACyear {2005}}%
}]{%
poldrack2005neural}
\APACinsertmetastar {%
poldrack2005neural}%
\begin{APACrefauthors}%
Poldrack, R\BPBI A.%
, Sabb, F\BPBI W.%
, Foerde, K.%
, Tom, S\BPBI M.%
, Asarnow, R\BPBI F.%
, Bookheimer, S\BPBI Y.%
\BCBL {}\ \BBA {} Knowlton, B\BPBI J.%
\end{APACrefauthors}%
\unskip\
\newblock
\APACrefYearMonthDay{2005}{}{}.
\newblock
{\BBOQ}\APACrefatitle {The neural correlates of motor skill automaticity} {The
  neural correlates of motor skill automaticity}.{\BBCQ}
\newblock
\APACjournalVolNumPages{J. Neurosci.}{25}{22}{5356--5364}.
\PrintBackRefs{\CurrentBib}

\bibitem [\protect \citeauthoryear {%
Pyle%
\ \BBA {} Rosenbaum%
}{%
Pyle%
\ \BBA {} Rosenbaum%
}{%
{\protect \APACyear {2017}}%
}]{%
pyle2017spatiotemporal}
\APACinsertmetastar {%
pyle2017spatiotemporal}%
\begin{APACrefauthors}%
Pyle, R.%
\BCBT {}\ \BBA {} Rosenbaum, R.%
\end{APACrefauthors}%
\unskip\
\newblock
\APACrefYearMonthDay{2017}{}{}.
\newblock
{\BBOQ}\APACrefatitle {Spatiotemporal dynamics and reliable computations in
  recurrent spiking neural networks} {Spatiotemporal dynamics and reliable
  computations in recurrent spiking neural networks}.{\BBCQ}
\newblock
\APACjournalVolNumPages{Phys. Rev. Lett.}{118}{1}{018103}.
\PrintBackRefs{\CurrentBib}

\bibitem [\protect \citeauthoryear {%
Russo%
\ \protect \BOthers {.}}{%
Russo%
\ \protect \BOthers {.}}{%
{\protect \APACyear {2018}}%
}]{%
russo2018motor}
\APACinsertmetastar {%
russo2018motor}%
\begin{APACrefauthors}%
Russo, A\BPBI A.%
, Bittner, S\BPBI R.%
, Perkins, S\BPBI M.%
, Seely, J\BPBI S.%
, London, B\BPBI M.%
, Lara, A\BPBI H.%
\BDBL {}others%
\end{APACrefauthors}%
\unskip\
\newblock
\APACrefYearMonthDay{2018}{}{}.
\newblock
{\BBOQ}\APACrefatitle {Motor cortex embeds muscle-like commands in an untangled
  population response} {Motor cortex embeds muscle-like commands in an
  untangled population response}.{\BBCQ}
\newblock
\APACjournalVolNumPages{Neuron}{97}{4}{953--966}.
\PrintBackRefs{\CurrentBib}

\bibitem [\protect \citeauthoryear {%
Seung%
}{%
Seung%
}{%
{\protect \APACyear {2003}}%
}]{%
seung2003learning}
\APACinsertmetastar {%
seung2003learning}%
\begin{APACrefauthors}%
Seung, H\BPBI S.%
\end{APACrefauthors}%
\unskip\
\newblock
\APACrefYearMonthDay{2003}{}{}.
\newblock
{\BBOQ}\APACrefatitle {Learning in spiking neural networks by reinforcement of
  stochastic synaptic transmission} {Learning in spiking neural networks by
  reinforcement of stochastic synaptic transmission}.{\BBCQ}
\newblock
\APACjournalVolNumPages{Neuron}{40}{6}{1063--1073}.
\PrintBackRefs{\CurrentBib}

\bibitem [\protect \citeauthoryear {%
Shenoy%
, Sahani%
\BCBL {}\ \BBA {} Churchland%
}{%
Shenoy%
\ \protect \BOthers {.}}{%
{\protect \APACyear {2013}}%
}]{%
Shenoy2013}
\APACinsertmetastar {%
Shenoy2013}%
\begin{APACrefauthors}%
Shenoy, K\BPBI V.%
, Sahani, M.%
\BCBL {}\ \BBA {} Churchland, M\BPBI M.%
\end{APACrefauthors}%
\unskip\
\newblock
\APACrefYearMonthDay{2013}{}{}.
\newblock
{\BBOQ}\APACrefatitle {{Cortical control of arm movements: a dynamical systems
  perspective.}} {{Cortical control of arm movements: a dynamical systems
  perspective.}}{\BBCQ}
\newblock
\APACjournalVolNumPages{Annu. Rev. Neurosci.}{36}{}{337--59}.
\PrintBackRefs{\CurrentBib}

\bibitem [\protect \citeauthoryear {%
Sompolinsky%
, Crisanti%
\BCBL {}\ \BBA {} Sommers%
}{%
Sompolinsky%
\ \protect \BOthers {.}}{%
{\protect \APACyear {1988}}%
}]{%
Sompolinsky1988}
\APACinsertmetastar {%
Sompolinsky1988}%
\begin{APACrefauthors}%
Sompolinsky, H.%
, Crisanti, a.%
\BCBL {}\ \BBA {} Sommers, H\BPBI J.%
\end{APACrefauthors}%
\unskip\
\newblock
\APACrefYearMonthDay{1988}{}{}.
\newblock
{\BBOQ}\APACrefatitle {{Chaos in random neural networks}} {{Chaos in random
  neural networks}}.{\BBCQ}
\newblock
\APACjournalVolNumPages{Phys. Rev. Lett.}{61}{3}{259--262}.
\PrintBackRefs{\CurrentBib}

\bibitem [\protect \citeauthoryear {%
Sussillo%
}{%
Sussillo%
}{%
{\protect \APACyear {2014}}%
}]{%
Sussillo2014}
\APACinsertmetastar {%
Sussillo2014}%
\begin{APACrefauthors}%
Sussillo, D.%
\end{APACrefauthors}%
\unskip\
\newblock
\APACrefYearMonthDay{2014}{}{}.
\newblock
{\BBOQ}\APACrefatitle {{Neural circuits as computational dynamical systems}}
  {{Neural circuits as computational dynamical systems}}.{\BBCQ}
\newblock
\APACjournalVolNumPages{Curr. Opin. Neurobiol.}{25}{}{156--163}.
\PrintBackRefs{\CurrentBib}

\bibitem [\protect \citeauthoryear {%
Sussillo%
\ \BBA {} Abbott%
}{%
Sussillo%
\ \BBA {} Abbott%
}{%
{\protect \APACyear {2009}}%
}]{%
sussillo2009generating}
\APACinsertmetastar {%
sussillo2009generating}%
\begin{APACrefauthors}%
Sussillo, D.%
\BCBT {}\ \BBA {} Abbott, L\BPBI F.%
\end{APACrefauthors}%
\unskip\
\newblock
\APACrefYearMonthDay{2009}{}{}.
\newblock
{\BBOQ}\APACrefatitle {Generating coherent patterns of activity from chaotic
  neural networks} {Generating coherent patterns of activity from chaotic
  neural networks}.{\BBCQ}
\newblock
\APACjournalVolNumPages{Neuron}{63}{4}{544--557}.
\PrintBackRefs{\CurrentBib}

\bibitem [\protect \citeauthoryear {%
Sussillo%
, Churchland%
, Kaufman%
\BCBL {}\ \BBA {} Shenoy%
}{%
Sussillo%
\ \protect \BOthers {.}}{%
{\protect \APACyear {2013}}%
}]{%
Sussillo2013}
\APACinsertmetastar {%
Sussillo2013}%
\begin{APACrefauthors}%
Sussillo, D.%
, Churchland, M\BPBI M.%
, Kaufman, M\BPBI T.%
\BCBL {}\ \BBA {} Shenoy, K\BPBI V.%
\end{APACrefauthors}%
\unskip\
\newblock
\APACrefYearMonthDay{2013}{}{}.
\newblock
{\BBOQ}\APACrefatitle {{A neural network that finds naturalistic solutions for
  the production of muscle activity}} {{A neural network that finds
  naturalistic solutions for the production of muscle activity}}.{\BBCQ}
\newblock
\APACjournalVolNumPages{Nat. Neurosci.}{18}{7}{}.
\PrintBackRefs{\CurrentBib}

\bibitem [\protect \citeauthoryear {%
Tang%
\ \protect \BOthers {.}}{%
Tang%
\ \protect \BOthers {.}}{%
{\protect \APACyear {2009}}%
}]{%
Tang2009}
\APACinsertmetastar {%
Tang2009}%
\begin{APACrefauthors}%
Tang, C\BPBI C.%
, Root, D\BPBI H.%
, Duke, D\BPBI C.%
, Zhu, Y.%
, Teixeria, K.%
, Ma, S.%
\BDBL {}West, M\BPBI O.%
\end{APACrefauthors}%
\unskip\
\newblock
\APACrefYearMonthDay{2009}{}{}.
\newblock
{\BBOQ}\APACrefatitle {{Decreased Firing of Striatal Neurons Related to Licking
  during Acquisition and Overtraining of a Licking Task}} {{Decreased Firing of
  Striatal Neurons Related to Licking during Acquisition and Overtraining of a
  Licking Task}}.{\BBCQ}
\newblock
\APACjournalVolNumPages{J. Neurosci.}{29}{44}{13952--13961}.
\PrintBackRefs{\CurrentBib}

\bibitem [\protect \citeauthoryear {%
Toledo-Su{\'a}rez%
, Duarte%
\BCBL {}\ \BBA {} Morrison%
}{%
Toledo-Su{\'a}rez%
\ \protect \BOthers {.}}{%
{\protect \APACyear {2014}}%
}]{%
toledo2014liquid}
\APACinsertmetastar {%
toledo2014liquid}%
\begin{APACrefauthors}%
Toledo-Su{\'a}rez, C.%
, Duarte, R.%
\BCBL {}\ \BBA {} Morrison, A.%
\end{APACrefauthors}%
\unskip\
\newblock
\APACrefYearMonthDay{2014}{}{}.
\newblock
{\BBOQ}\APACrefatitle {Liquid computing on and off the edge of chaos with a
  striatal microcircuit} {Liquid computing on and off the edge of chaos with a
  striatal microcircuit}.{\BBCQ}
\newblock
\APACjournalVolNumPages{Frontiers in computational neuroscience}{8}{}{130}.
\PrintBackRefs{\CurrentBib}

\bibitem [\protect \citeauthoryear {%
Turner%
\ \BBA {} Desmurget%
}{%
Turner%
\ \BBA {} Desmurget%
}{%
{\protect \APACyear {2010}}%
}]{%
Turner2010}
\APACinsertmetastar {%
Turner2010}%
\begin{APACrefauthors}%
Turner, R\BPBI S.%
\BCBT {}\ \BBA {} Desmurget, M.%
\end{APACrefauthors}%
\unskip\
\newblock
\APACrefYearMonthDay{2010}{}{}.
\newblock
{\BBOQ}\APACrefatitle {{Basal ganglia contributions to motor control: a
  vigorous tutor.}} {{Basal ganglia contributions to motor control: a vigorous
  tutor.}}{\BBCQ}
\newblock
\APACjournalVolNumPages{Curr. Opin. Neurobiol.}{20}{6}{704--16}.
\PrintBackRefs{\CurrentBib}

\bibitem [\protect \citeauthoryear {%
Vincent-Lamarre%
, Lajoie%
\BCBL {}\ \BBA {} Thivierge%
}{%
Vincent-Lamarre%
\ \protect \BOthers {.}}{%
{\protect \APACyear {2016}}%
}]{%
vincent2016driving}
\APACinsertmetastar {%
vincent2016driving}%
\begin{APACrefauthors}%
Vincent-Lamarre, P.%
, Lajoie, G.%
\BCBL {}\ \BBA {} Thivierge, J\BHBI P.%
\end{APACrefauthors}%
\unskip\
\newblock
\APACrefYearMonthDay{2016}{}{}.
\newblock
{\BBOQ}\APACrefatitle {Driving reservoir models with oscillations: a solution
  to the extreme structural sensitivity of chaotic networks} {Driving reservoir
  models with oscillations: a solution to the extreme structural sensitivity of
  chaotic networks}.{\BBCQ}
\newblock
\APACjournalVolNumPages{J. Comput. Neurosci.}{41}{3}{305--322}.
\PrintBackRefs{\CurrentBib}

\bibitem [\protect \citeauthoryear {%
Xie%
\ \BBA {} Seung%
}{%
Xie%
\ \BBA {} Seung%
}{%
{\protect \APACyear {2004}}%
}]{%
xie2004learning}
\APACinsertmetastar {%
xie2004learning}%
\begin{APACrefauthors}%
Xie, X.%
\BCBT {}\ \BBA {} Seung, H\BPBI S.%
\end{APACrefauthors}%
\unskip\
\newblock
\APACrefYearMonthDay{2004}{}{}.
\newblock
{\BBOQ}\APACrefatitle {Learning in neural networks by reinforcement of
  irregular spiking} {Learning in neural networks by reinforcement of irregular
  spiking}.{\BBCQ}
\newblock
\APACjournalVolNumPages{Physical Review E}{69}{4}{041909}.
\PrintBackRefs{\CurrentBib}

\end{thebibliography}

\end{document}